\documentclass{aa}  

\usepackage{CJKutf8}
\usepackage{amsmath}
\usepackage{amssymb}
\usepackage{amsfonts}
\usepackage{graphicx}
\usepackage{wasysym}
\usepackage{multirow}
\usepackage{mdframed}
\usepackage[T1]{fontenc}

\newcommand{\meth}{\mbox{CH$_3$OH}}

\newcommand{\fmh}{\mbox{H$_2$CO}}
\newcommand{\httco}{\mbox{H$_2$$^{13}$CO}}
\newcommand{\mthc}{\mbox{CH$_3$CN}}
\newcommand{\cyacet}{\mbox{HC$_3$N}}
\newcommand{\cctht}{\mbox{c-C$_3$H$_2$}}

\newcommand{\hcop}{\mbox{HCO$^+$}}

\newcommand{\kms}{\mbox{km\,s$^{-1}$}}

\newcommand{\sqc}{\mbox{cm$^{-2}$}}
\newcommand{\cc}{\mbox{cm$^{-3}$}}

\newcommand{\msol}{\mbox{$M_\odot$}}

\newcommand{\mujypbm}{\mbox{$\mu$Jy\,beam$^{-1}$}}
\newcommand{\mjypbm}{\mbox{mJy\,beam$^{-1}$}}

\newcommand{\ctw}{the 20~\kms{} cloud}
\newcommand{\cfi}{the 50~\kms{} cloud}

\defcitealias{lu2020}{Paper~I}
\defcitealias{lu2021}{Paper~II}

\usepackage{graphicx}
\usepackage[breaklinks,colorlinks,urlcolor=blue,citecolor=blue,linkcolor=blue]{hyperref}
\usepackage{multirow}

\usepackage{txfonts}

\begin{document}

\title{ALMA observations of massive clouds in the central molecular zone: slim filaments tracing parsec-scale shocks}

   \titlerunning{Gas Filaments in CMZ Clouds}

   \author{Kai Yang\inst{1,2,3}
          \and
          Xing Lu\inst{4}
          \and
          Yichen Zhang\inst{1}
          \and
          Xunchuan Liu\inst{4}
          \and
          Adam Ginsburg\inst{5}
          \and
          Hauyu Baobab Liu\inst{6,7}
          \and
          Yu Cheng\inst{8}
          \and
          Siyi Feng\inst{9}
          \and
          Tie Liu\inst{4}
          \and
          Qizhou Zhang\inst{10}
          \and
          Elisabeth A.C. Mills\inst{11}
          \and
          Daniel L.\ Walker\inst{12}
          \and
          Shu-ichiro Inutsuka\inst{13}
          \and
          Cara Battersby\inst{14}
          \and
          Steven N. Longmore\inst{15,16}
          \and
          Xindi Tang\inst{17,18,19,20}
          \and
          Jens Kauffmann\inst{21}
          \and
          Qilao Gu\inst{4}
          \and
          Shanghuo Li\inst{2,3}
          \and
          Qiuyi Luo\inst{4}
          \and
          J.~M.~Diederik~Kruijssen\inst{22,16}
          \and
          Thushara Pillai\inst{21}
          \and
          Hai-Hua Qiao\inst{23,24,4}
          \and
          Keping Qiu\inst{2,3}
          \and
          Zhiqiang Shen\inst{4}
          }

   \institute{Department of Astronomy, School of Physics and Astronomy, Shanghai Jiao Tong University, 800 Dongchuan Rd., Minhang, Shanghai 200240, China\\
              \email{kyang2146@sjtu.edu.cn, yichen.zhang@sjtu.edu.cn}
         \and
             School of Astronomy and Space Science, Nanjing University, 163 Xianlin Avenue, Nanjing 210023, China
         \and
             Key Laboratory of Modern Astronomy and Astrophysics (Nanjing University), Ministry of Education, Nanjing 210023, China
         \and
             Shanghai Astronomical Observatory, Chinese Academy of Sciences, 80 Nandan Road, Shanghai 200030, China \\
             \email{xinglu@shao.ac.cn}
         \and
             Department of Astronomy, University of Florida, P.O.\ Box 112055, Gainesville, FL 32611, USA
         \and
             Department of Physics, National Sun Yat-Sen University, No.\ 70, Lien-Hai Road, Kaohsiung City 80424, Taiwan
         \and
             Center of Astronomy and Gravitation, National Taiwan Normal University, Taipei 116, Taiwan
         \and
             National Astronomical Observatory of Japan, 2-21-1 Osawa, Mitaka, Tokyo 181-8588, Japan
         \and
             Department of Astronomy, Xiamen University, Zengcuo’an West Road, Xiamen, 361005, China
         \and
             Center for Astrophysics | Harvard \& Smithsonian, 60 Garden Street, Cambridge, MA 02138, USA
         \and
             Department of Physics and Astronomy, University of Kansas, 1251 Wescoe Hall Drive, Lawrence, KS 66045, USA
         \and
             UK ALMA Regional Centre Node, Jodrell Bank Centre for Astrophysics, The University of Manchester, Manchester M13 9PL, UK
         \and
             Department of Physics, Graduate School of Science, Nagoya University, Furo-cho, Chikusa-ku, Nagoya 464-8602, Japan
         \and
             Department of Physics, University of Connecticut, 196A Auditorium Road, Storrs, CT 06269, USA
         \and
             Astrophysics Research Institute, Liverpool John Moores University, IC2, Liverpool Science Park, 146 Brownlow Hill, Liverpool, L3 5RF, United Kingdom
         \and
             Cosmic Origins Of Life (COOL) Research DAO, Munich, Germany
         \and
             Xinjiang Astronomical Observatory, 150 Science 1-Street, Urumqi, Xinjiang 830011, China
         \and
             University of Chinese Academy of Sciences, Beijing, 100080, China
         \and
             Key Laboratory of Radio Astronomy, Chinese Academy of Sciences, Urumqi, 830011, China
         \and
             Xinjiang Key Laboratory of Radio Astrophysics,Urumqi, 830011, China
         \and
             Haystack Observatory, Massachusetts Institute of Technology, 99 Millstone Road, Westford, MA 01886, USA
         \and
             Chair of Remote Sensing Technology, School of Engineering and Design, Department of Aerospace and Geodesy, Technical University of Munich, Arcisstra\ss e 21, D-80333 Munich, Germany
         \and
             National Time Service Center, Chinese Academy of Sciences, Xi'An, Shaanxi 710600, China
         \and
             Key Laboratory of Time Reference and Applications, Chinese Academy of Sciences, China
             }
    \authorrunning{Yang, K.\ et al.}
   \date{Received XXX; accepted XXX}

\abstract
{The central molecular zone (CMZ) of our Galaxy exhibits widespread emission from SiO and various complex organic molecules (COMs), yet the exact origin of such emission is uncertain.
Here we report the discovery of a unique class of long ($>$0.5~pc) and narrow ($<$0.03~pc) filaments in the emission of SiO 5--4 and eight additional molecular lines, including several COMs, in our ALMA 1.3~mm spectral line observations toward two massive molecular clouds in the CMZ, which we name as slim filaments.
However, these filaments are not detected in the 1.3~mm continuum at the 5$\sigma$ level. 
Their line-of-sight velocities are coherent and inconsistent with being outflows. The column densities and relative abundances of the detected molecules are statistically similar to those in protostellar outflows but different from those in dense cores within the same clouds. Turbulent pressure in these filaments dominates over self gravity and leads to hydrostatic inequilibrium, indicating that they are a different class of objects than the dense gas filaments in dynamical equilibrium ubiquitously found in nearby molecular clouds.
We argue that these newly detected slim filaments are associated with parsec-scale shocks, likely arising from dynamic interactions between shock waves and molecular clouds. The dissipation of the slim filaments may replenish SiO and COMs in the interstellar medium and lead to their widespread emission in the CMZ.}


   \maketitle
%

\section{Introduction}\label{sec:intro}

The central molecular zone (CMZ), usually referring to the inner 500~pc of the Galaxy, is a gas reservoir of total mass of several times $10^7$~\msol{} and mean density at $10^4$~\cc{} \citep{morris1996,ferriere2007,longmore2013a} yet with unexpectedly inefficient star formation \citep{longmore2013a,kruijssen2014,barnes2017,2023ASPC..534...83H}. Prominent parsec-scale shocks have been suggested to be widespread in the CMZ, manifested by the measured molecular line widths of 5--10~\kms{}, a factor of $\gtrsim$10 broader than those measured toward nearby clouds at the same scale \citep{baobab2013,henshaw2016a}, as well as the ubiquitous SiO emission in the cloud scale \citep{martinpintado1997,baobab2013,minh2015}.  
The shocks are suggested to heat the gas, leading to the de-coupling of gas and dust temperatures in the clouds \citep{ao2013,mills2013,ginsburg2016,immer2016,krieger2017,lu2017}, and to be related to the unique ``hot-core like'' chemistry with widespread complex organic molecules (COMs) in the cloud scale \citep{martinpintado2001,requenatorres2006,requenatorres2008,menten2009}.

The origin of the parsec-scale shocks in the CMZ is unclear. Similar parsec-scale shocks have been discussed toward massive star forming clouds in the Galactic disk \citep{jimenezserra2010,jimenezserra2014,nguyenluong2013,cosentino2018}, which are often attributed to collisions between clouds. Dynamic interactions such as collisions, shear motions, or inflow along the bar may be more frequent for clouds in the CMZ given the high volume densities of clouds \citep{kruijssen2014,kruijssen2019a,li2020,inutsuka2021}.

Recent high resolution observations using the Atacama Large Millimeter/submillimeter Array (ALMA) have detected filamentary molecular gas structures in the CMZ clouds. \citet{bally2014} discovered molecular absorption filaments in \hcop{} in the massive cloud G0.253+0.016, and proposed that the broad-line absorption filaments (with line widths $>$ 20 \kms{}) could be foreground magnetic structures and the narrow-line absorption filaments (with line widths $<$ 20 \kms{}) may trace optically thick gas on the front expanding surface of this cloud. \citet{wallace2022} identified CO filaments running parallel to the Galactic plane in the Sgr~E cloud and argued that the gravitational influence of the Galactic bar is responsible for these filaments by stretching molecular gas in this region. 
\citet{2022MNRAS.509.4758H} identified an arc-shaped HNCO emission feature in G0.253+0.016, which is believed to result from a bubble swept up by winds or stellar feedback driven by star formation. In addition, a bubble-shaped HNCO emission feature M0.8$-$0.2 was reported by \citet{2024A&A...691A..70N}, which is thought to be the outcome of a high-energy hypernova explosion.
However, none of these filaments were thought to be directly related to parsec-scale shocks. Meanwhile, in nearby star-forming regions, interferometric observations have identified shock-related filamentary structures using shock tracers, particularly SiO \citep[e.g.][]{2019ApJ...881L..42C, 2022A&A...667A...6C,desimone2022}. 

Using emissions from nine spectral lines at 1.3~mm, observed with ALMA toward two massive clouds in the CMZ, we have discovered a population of narrow ($<$0.03~pc) filamentary structures, which are not visible in dust emission, termed ``slim filaments''. These slim filaments are likely related to parsec-scale shocks, which could be indicative of active dynamic interactions within the clouds. We adopt a distance of 8.1~kpc to the CMZ \citep{reid2019}.

\section{Observations}\label{sec:obs}

\begin{table}
\caption{Spectral lines detected in the slim filaments.}  \label{tab_lines}
\begin{tabular}{l l c r c}
\hline\hline
Molecule & Transition & Rest frequency & $E_{u}/k$ & $n_{\rm crit}^{(a)}$\\
 &  & (MHz) & (K) & (cm$^{-3}$)\\
\hline
 SiO & 5--4 & 217104.98 & 31.3 & 2.6$\times$10$^{6}$ \\ 
 \cctht{} & 6$_{1,6}$--5$_{0,5}$ & 217822.15 & 38.6 & 4.5$\times$10$^{7}$ \\ 
 \fmh{} & 3$_{0,3}$--2$_{0,2}$ & 218222.19 & 21.0 & 3.4$\times$10$^{6}$ \\ 
 \cyacet{} & 24--23 & 218324.72 & 131.0 & 1.7$\times$10$^{7}$ \\ 
 \meth{} & 4$_{2,2}$--3$_{1,2}$ & 218440.06 & 45.5 & 7.8$\times$10$^{7}$ \\ 
 HNCO & 10$_{0,10}$--9$_{0,9}$ & 219798.27 & 58.0 & 7.6$\times$10$^{6}$ \\ 
 \httco{} & 3$_{1,2}$--2$_{1,1}$ & 219908.53 & 32.9 & 4.8$\times$10$^{6}$ \\
 SO & 6$_{5}$--5$_{4}$ & 219949.44 & 35.0 & 2.3$\times$10$^{6}$ \\ 
 \mthc{} & 12$_{1}$--11$_{1}$ & 220743.01 & 76.0 & 4.5$\times$10$^{6}$ \\ 
 & 12$_{0}$--11$_{0}$ & 220747.26 & 68.9 & 4.1$\times$10$^{6}$ \\
\hline\hline
\end{tabular}
\\
Notes. $^{\rm (a)}$ The critical density is calculated based on Einstein coefficients $A_{\rm ul}$ and collisional rate coefficients $C_{\rm ul}$ from the Leiden Atomic and Molecular Database \citep[LAMDA,][]{lamda2005}: via the approximation $n_{\rm crit}$ = $A_{\rm ul}$/$C_{\rm ul}$ \citep{shirley2015} at a temperature of $\sim$100~K. For the case of \httco{}, we give the critical density for the 3$_{1,2}$--2$_{1,1}$ transition of the main isotope.
\end{table}

The ALMA observations and data reduction were detailed in \citet[][hereafter \citetalias{lu2020}]{lu2020} and \citet[][hereafter \citetalias{lu2021}]{lu2021}, which have focused on 2000-au scale cores and protostellar outflows in the CMZ, respectively. Here, we summarize the key points. The observations toward four clouds in the CMZ were performed with the C40-3 and C40-5 configurations in 2017 April and July (project code: 2016.1.00243.S) and within frequency ranges of 217--221~GHz and 231--235~GHz. The sample includes \ctw{}, \cfi{}, cloud~e, and Sgr~C. The image cubes have been produced with the tclean task in CASA following the procedures outlined in \citetalias{lu2021}. The synthesized beam size of the images is on average 0\farcs{28}$\times$0\farcs{19} (equivalent to 2200~AU$\times$1500~AU) but slightly varies between lines. The maximum recoverable angular scale is 10\arcsec{} ($\sim$0.4~pc). The continuum rms measured in the emission-free regions without primary beam corrections is 40~\mujypbm, with a central frequency of 226~GHz. The spectral line rms is between 1.6--2.0~\mjypbm{} (0.8--1.0~K in brightness temperatures) per 0.976~MHz channel (corresponding to 1.35~\kms{} at 217.105~GHz).

\begin{figure*}
\centering
 \includegraphics[width=490pt]{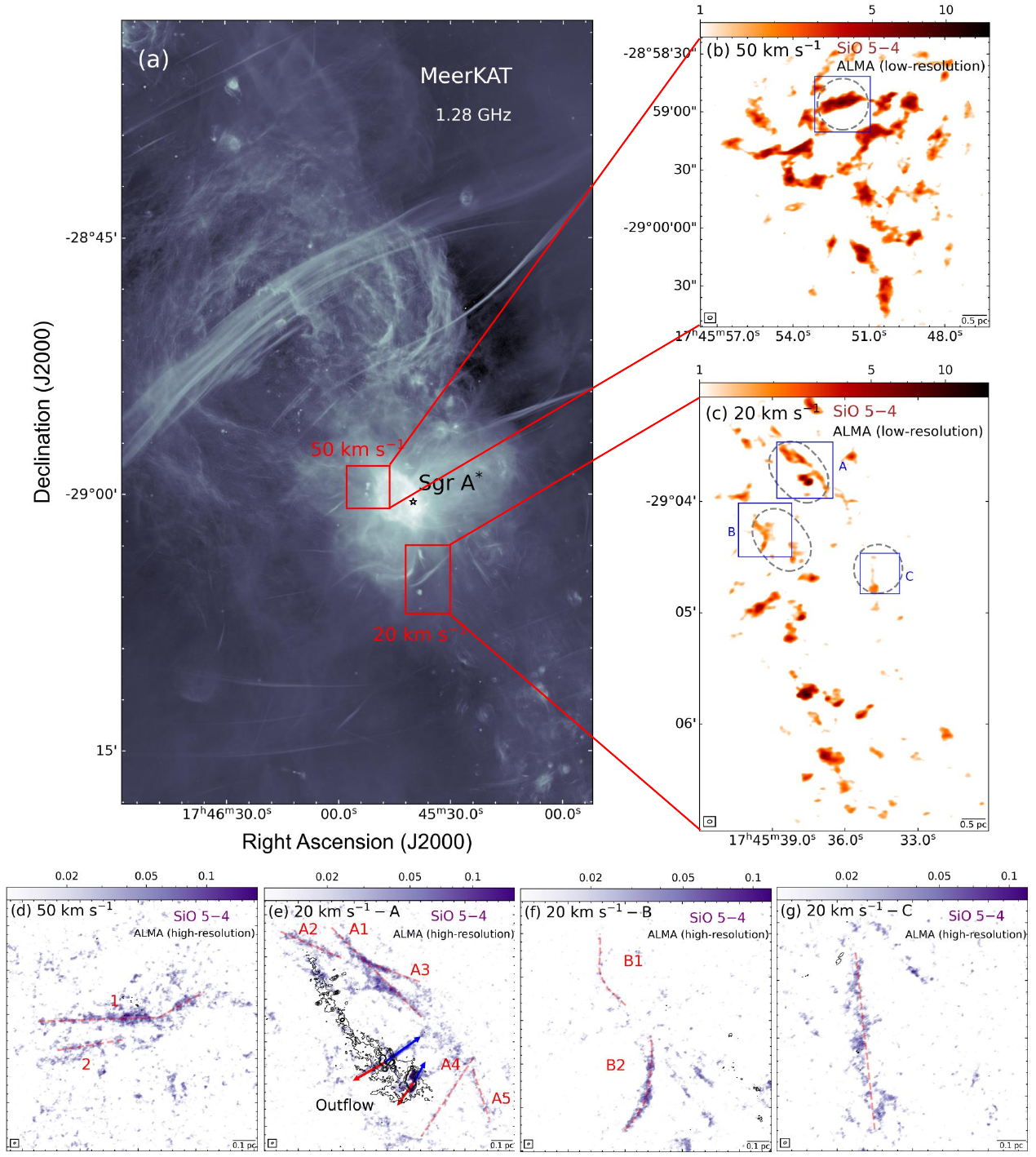}
 \caption{Slim filaments in the CMZ. Panel (a): MeerKAT 1.28~GHz radio emission of the Sgr~A region. The red boxes mark \ctw{} and \cfi{}. Panels (b)--(c): Integrated intensity maps of SiO\,5--4 in \ctw{} and \cfi{} from ALMA low-resolution ($\sim$1.9$^{\prime\prime}$) observations (project code: 2016.1.00875.S). The blue boxes mark zoom-in regions where slim filaments are detected. The dashed loops demonstrate the 50\% primary beam of our ALMA high-resolution ($\sim$0\farcs{23}) observation. Panels (d)--(g): SiO\,5--4 emission of filaments from our ALMA high-resolution observations, which are integrated in velocity ranges of [$-$20, 40] and [25, 75] \kms for \ctw{} and \cfi{} respectively. The pink dashed lines illustrate the identified slim filaments. The black contours present the ALMA 1.3 mm continuum emission at levels of [5, 25, 45] $\times$ 40~\mujypbm.}
 \label{fig:zoomin}
\end{figure*}

\section{Results}\label{sec:res}

\subsection{Identification of the slim filaments} \label{subsubsec:identi}

\citetalias{lu2021} identified 43 protostellar outflows from emissions of six potential shock tracers (SiO, SO, HNCO, \fmh{}, \cyacet{}, and \meth{}, with transitions listed in \autoref{tab_lines}). Interestingly, apart from these outflows, filamentary structures not associated with any dust emission are found in three regions in \ctw{} and one region in \cfi{}. In this letter, we focused on the SiO 5--4 transition that has been suggested to primarily trace shock activities, including not only outflow-associated fast shocks (e.g., \citealt{2007prpl.conf..245A}; \citealt{2007ApJ...654..361Q}; \citealt{2011A&A...526L...2L}; \citealt{2024ApJ...960...48T}), but also slow shocks (with emissions mostly within $\pm$5~\kms{} of the cloud $V_{\rm LSR}$, e.g., \citealt{2007A&A...476.1243M}; \citealt{2014A&A...570A...1D}; \citealt{2016A&A...586A.149C}; \citealt{2016A&A...595A.122L}; \citealt{2016ApJ...824...99M}; \citealt{2024A&A...684A.140Y}).

As shown in panel (a) of \autoref{fig:zoomin}, a MeerKAT continuum map \citep{heywood2022} illustrates the positions of the two filament-detected clouds relative to Sgr~A$^{\star}$. Panels (b) and (c) present the integrated intensity maps of the SiO 5--4 line toward the clouds, obtained from ALMA low-resolution ($\sim$1.9$^{\prime\prime}$) observations (project code: 2016.1.00875.S). Panels (d)--(g) display zoomed-in views of the filamentary structures captured with our ALMA high-resolution observations.

We identified the slim filaments through visual inspection by the following criteria:
the filaments should have spatially continuous SiO 5--4 integrated line emission with a signal-to-noise ratio greater than 3, a coherent velocity structure in the position-velocity diagram, and an aspect ratio greater than 10. The skeletons, lengths, widths, and velocity structures of the filaments are detailed in Section \ref{subsec:basic_para}. In \cfi{}, we detected two slim filaments oriented in the west-east direction. In the A region of \ctw{}, \citetalias{lu2021} identified two bipolar outflows. Additionally, a filamentary structure with several branches was identified north of the outflows, and two straight filaments were found to the west of the outflows. In the B and C regions, we found two curved filaments and a north-south oriented filament, respectively. Notably, these filaments are not detected in the dust emission at the 5$\sigma$ level of 0.2~\mjypbm{}. As shown in panels (b) \& (c) of \autoref{fig:zoomin}, similar filamentary emissions were also found in the low-resolution ALMA SiO images, suggesting that the slim filaments are unlikely to be a result of spatial filtering.

\begin{figure*}
\centering
\includegraphics[width=0.9\textwidth]{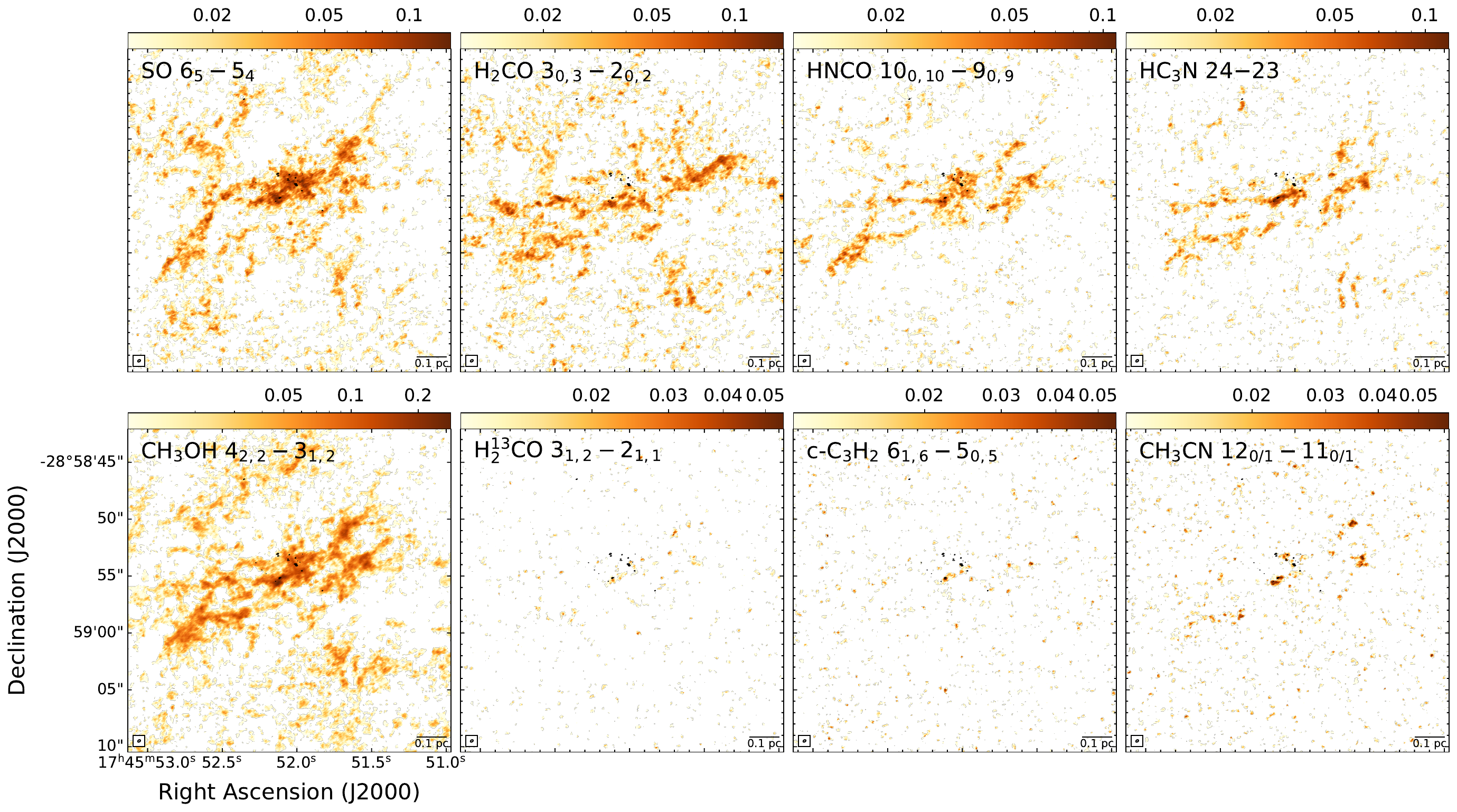}
\caption{Integrated maps of molecules in \autoref{tab_lines} for the slim filaments in the \cfi. The black contours present the ALMA 1.3 mm continuum emission at levels of [5, 25, 45] $\times$ 40~\mujypbm.}
\label{fig:m0_d_app}
\end{figure*}

The integrated intensity maps of other potential shock tracers in \cfi{} are presented in Figures~\ref{fig:m0_d_app}, where these filaments are also detected in the emissions of these tracers. The integrated maps of these molecular lines toward slim filaments in \ctw{} are shown in \autoref{app_sec:fila}. In some cases, the filaments are even marginally detected in the emission of \mthc{} 12$_{0/1}$--11$_{0/1}$, \httco{} 3$_{1,2}$--2$_{1,1}$, and \cctht{} 6$_{1,6}$--5$_{0,5}$, which are usually detected in star forming dense cores.

Considering their slim morphology and unique properties (e.g., high velocity dispersions, see Section \ref{subsec:basic_para}), these filaments represent distinctive structures within the CMZ that have not been seen elsewhere in the Galaxy. We note that the identified slim filaments are by no means a complete sample of such objects in the surveyed clouds. A few ambiguous candidates may exist in cloud~e and Sgr~C. \cite{2024arXiv241009253C} identified morphologically similar filaments in Sgr~C using the JWST-NIRCam H$_{2}$ image, which might trace shocked molecular gas as well. However, in this letter, we intended to report the first detection and characterization of such objects, and therefore only focused on the most robust candidates.

\subsection{Characteristics of the slim filaments}\label{subsec:basic_para}

\subsubsection{Widths}\label{subsec:disc_width}
\begin{figure}
\centering
 \includegraphics[width=0.4\textwidth]{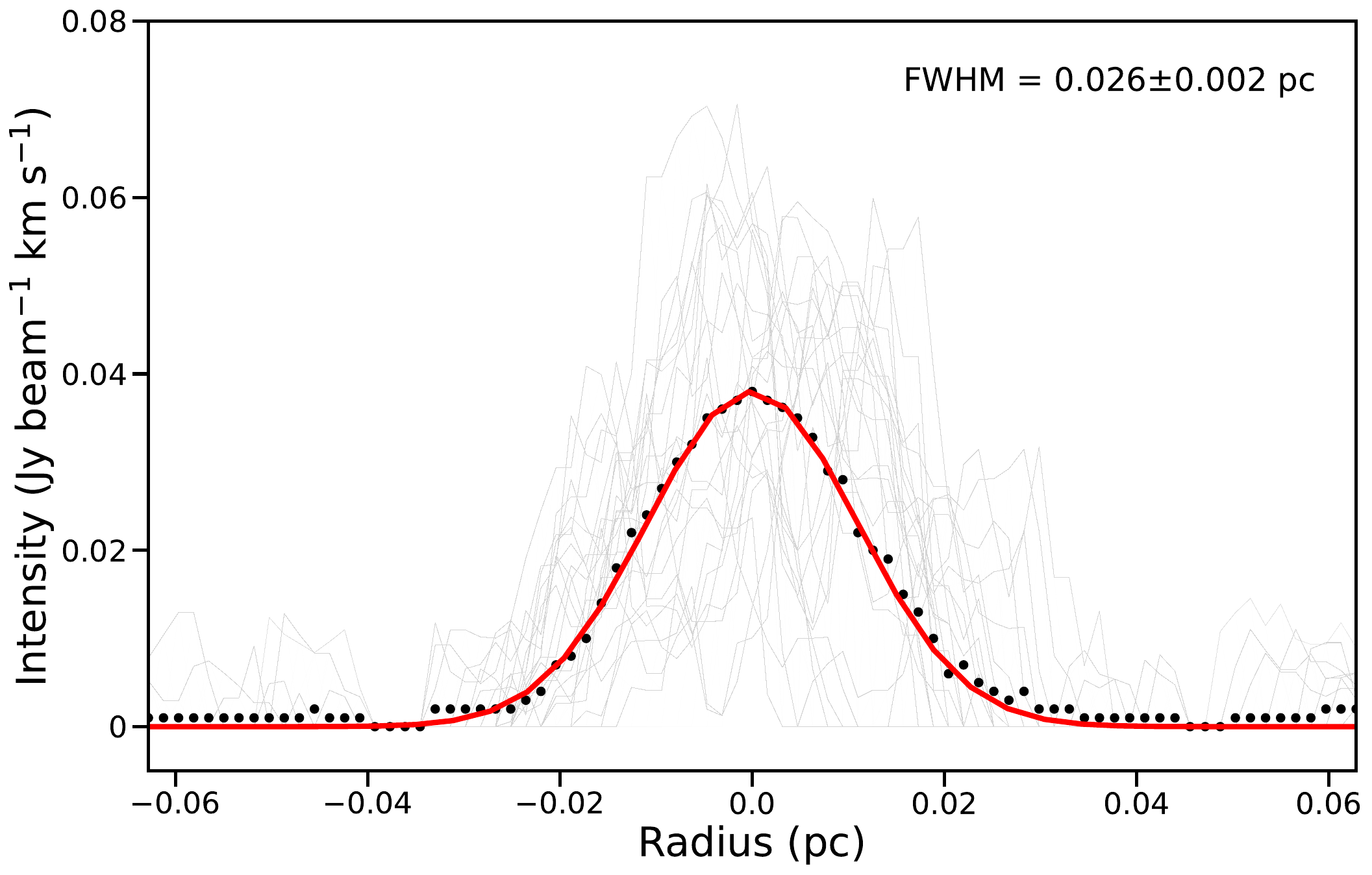}
 \caption{Mean radial intensity profile perpendicular to the filaments. Individual integrated intensity profiles are shown in gray, while the mean values are shown with black dots. The radius is the projected distance from the gas filament. The red solid line presents the best-fit result of Gaussian fitting.}
 \label{fig:widths}
\end{figure}

Based on the SiO 5--4 integrated intensity maps, we use the python-based package \texttt{FilFinder}\footnote{\url{https://github.com/e-koch/FilFinder}} \citep{2015MNRAS.452.3435K} to compute the filament skeletons (extraction processes are detailed in \autoref{app_sec:filfinder}). To determine the Full Width at Half Maximum (FWHM) of the filaments, we extract intensity profiles perpendicular to the skeletons at intervals of every 5 pixels (about one beam size) and perform Gaussian fits to the mean radial profile. The Gaussian function is expressed as:
\begin{equation}\label{fib_wid}
A(r) = A_{0}~{\rm exp}\left( \frac{-(r- \mu )^{2}}{2 \sigma^{2}_{\rm G}} \right) ,\\
\end{equation}
where $A(r)$ represents the profile amplitude at the radial distance $r$, $A_{0}$ is the amplitude, $\mu$ is the mean, and $\sigma_{\rm G}$ is the standard deviation. The best-fit FWHMs of \cfi{} and the 20~\kms{}-A, B, C regions are 0.025, 0.025, 0.026, and 0.028~pc, respectively. The overall best-fit FWHM for all the filaments is 0.026~pc, with the corresponding Gaussian curve shown as the red solid line in \autoref{fig:widths}. Considering the half-power beam width ($\rm FWHM_{\rm bm}$) of about 0.23$^{\prime\prime}$ ($\sim$ 0.009 pc), the beam-deconvolved FWHM can be estimated by FWHM$_{\rm decon}$ = $\sqrt{\rm FWHM^{2} - \rm FWHM_{\rm bm}^{2}}$ to be about 0.024~pc, approximately 3 times the beam size, which is narrower than the characteristic width of 0.1~pc toward dense gas filaments in nearby clouds \citep[e.g.,][]{2011A&A...529L...6A,2019A&A...621A..42A,2022A&A...667L...1A}. We will demonstrate in Section \ref{subsec:equi} that these slim filaments exhibit fundamentally different dynamic states compared to nearby filaments.

\subsubsection{Velocity dispersions}\label{subsec:linewidths}

The velocity dispersion, $\sigma_{v}$, is estimated as:
\begin{equation}\label{velo_disp}
\begin{split}
\sigma_{v} &= \sqrt{\sigma_{\rm nt,SiO}^{2}+c_{s}^{2}} \\
           &= \sqrt{\sigma_{\rm obs}^{2} - \frac{\Delta_{\rm ch}^{2}}{(2\sqrt{\rm 2ln2})^{2}} - \frac{k_{B}T}{\mu_{\rm SiO} m_{p}} + \frac{k_{B}T}{\mu_{p}m_{p}}}
\end{split}
\end{equation}
where $\sigma_{\rm nt}$ represents the non-thermal velocity dispersion, $c_{s}$ is the isothermal sound speed, $\Delta_{\rm ch}$ is channel width, $k_{B}$ is the Boltzmann constant, $T$ is gas temperature, $\mu_{\rm SiO}$ is the molecular weight of 44 for SiO, $\mu_{p}$ = 2.37 is the mean molecular weight per free particle \citep{2008A&A...487..993K}, and $m_{p}$ is the proton mass. The observed velocity dispersion, $\sigma_{\rm obs}$, is derived by fitting the averaged spectra (details provided in \autoref{appd_sec:widths}) with values of 5.7, 4.5, 4.5, and 3.8~\kms{} for \cfi{} and the 20 \kms{}-A, B, C regions, respectively. The typical gas temperature is approximately 70~K adopted from \citetalias{lu2021}. Finally, the velocity dispersions, $\sigma_{v}$, toward these four regions are found to be 5.6, 4.4, 4.4, and 3.7~\kms{} respectively, corresponding to FWHM line widths of 13.2, 10.4, 10.4, and 8.9~\kms.

\subsubsection{Velocity structures}\label{subsec:disc_pv}

\begin{figure*}
\centering
 \includegraphics[width=0.9\textwidth]{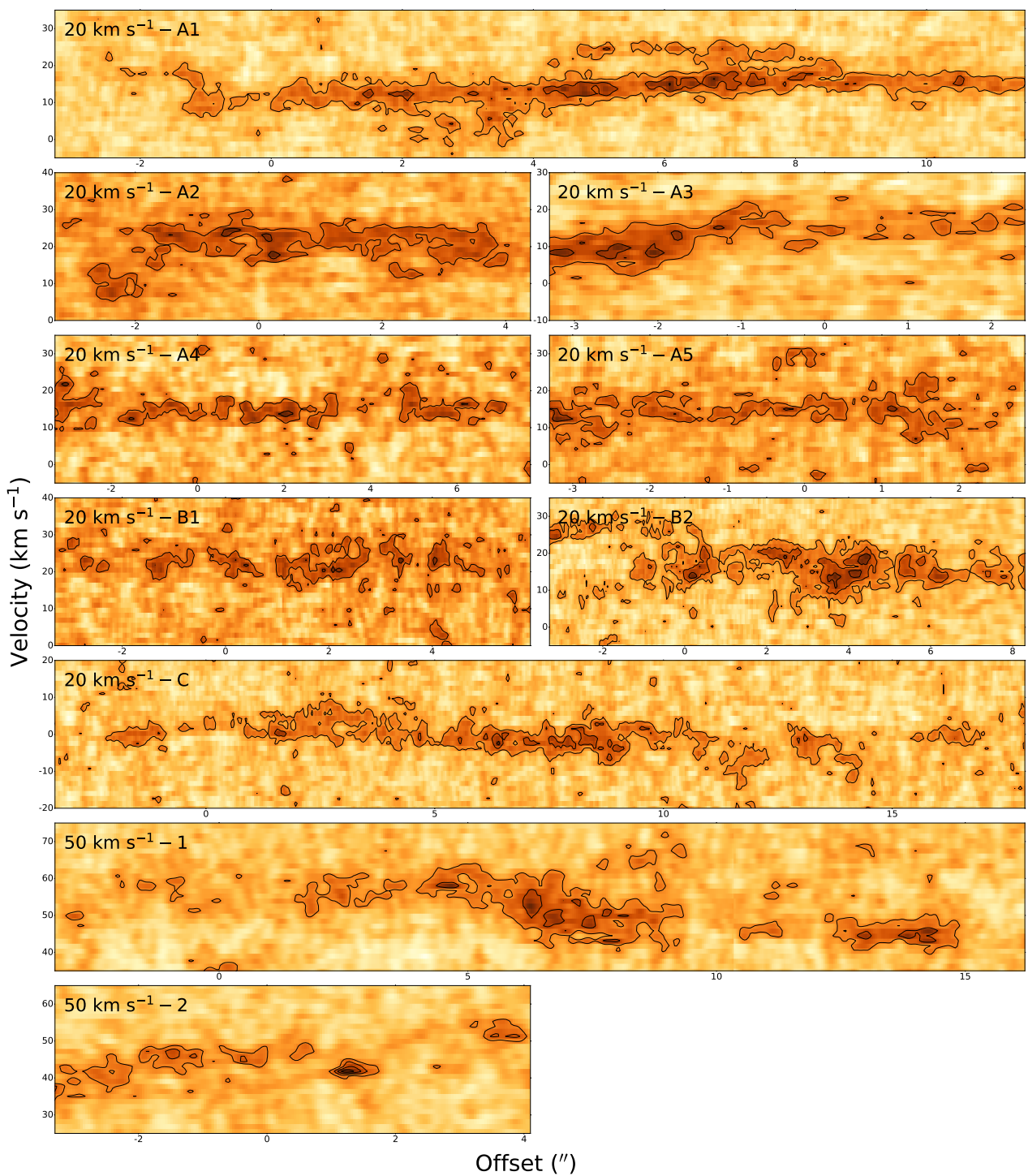}
 \caption{Position-velocity diagrams of SiO 5--4 emissions for the slim filaments along the dashed lines in \autoref{fig:zoomin}. The contour levels are (2 to 8 by steps of 2)$\times$$\sigma$, with $\sigma$ = 1.6~\mjypbm. The labels correspond to those in \autoref{fig:zoomin}.}
 \label{fig:pv}
\end{figure*}

The position-velocity (PV) diagrams for the slim filaments, shown in \autoref{fig:pv}, are derived along the dashed lines in \autoref{fig:zoomin}. Most of the slim filaments exhibit emissions around the systematic velocity, and their PV diagrams display consistent velocities along their major axes (with FWHM line widths of $\sim$10~\kms).

The SiO 5--4 transition was used to identify outflows in the CMZ in \citetalias{lu2021}.
However, as shown in \autoref{fig:pv}, the slim filaments do not show apparent velocity gradients along the major axes that are typically observed in outflows \cite[e.g.,][]{2001ApJ...554..132A,1996ApJ...459..638L}. Therefore, it is unlikely that they are outflows.

\subsection{Relative Abundances of the Molecules and Comparison to Outflows} \label{subsubsec:disc_column}

\begin{figure*}
\centering
\includegraphics[width=0.9\textwidth]{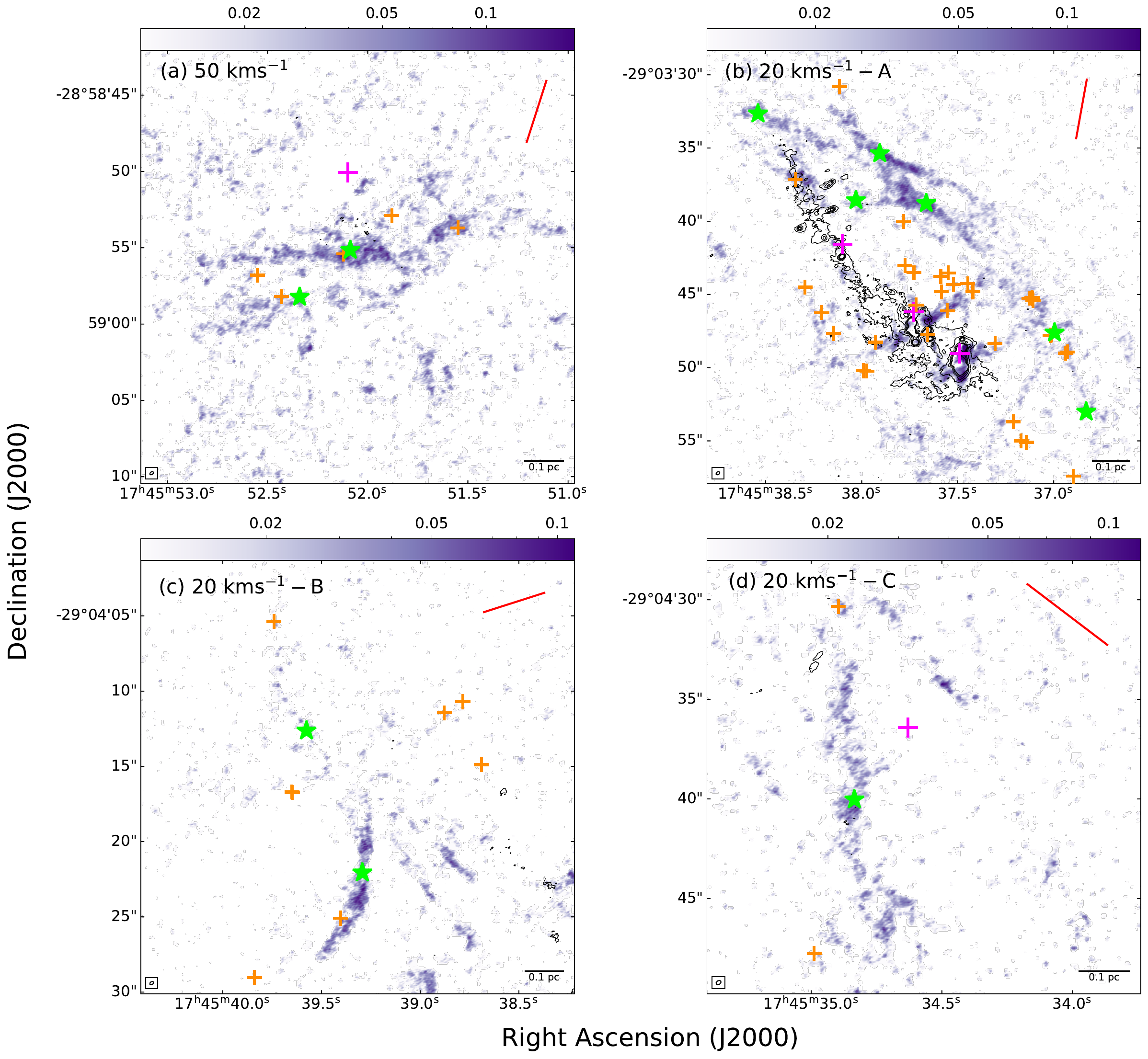}
\caption{Distributions of \meth{} and H$_{2}$O maser spots over the SiO\,5--4 integrated maps. The positions of \meth{} masers \citep{2011ApJ...739L..21P,cotton2016} are marked by orange crosses, while the positions of H$_{2}$O masers are indicated by magenta crosses obtained from \citet{lu2019b}. The green stars denote the reference positions selected for deriving the column densities of the molecules. The red lines in the upper-right corners illustrate the magnetic orientations \citep{2024ApJ...969..150P} with a resolution of 19.6$^{\prime\prime}$.}
\label{fig:maser}
\end{figure*}

Assuming local thermodynamic equilibrium (LTE) conditions and optically thin line emission, we estimated the column densities of the nine molecules (listed in \autoref{tab_lines}) at the SiO peak positions on the filament branches (shown as green stars in \autoref{fig:maser}). The method is detailed in \autoref{appd_sec:column}. Similar method has been used in \citepalias{lu2021} for calculating column densities of six molecules toward protostellar outflows in the CMZ. However, without detectable dust emission, we were unable to constrain the total molecular hydrogen column densities in the filaments. Consequently, we could only provide lower limits for molecular abundances and relative abundances between the molecules.

\begin{figure}
\centering
\includegraphics[width=0.48\textwidth]{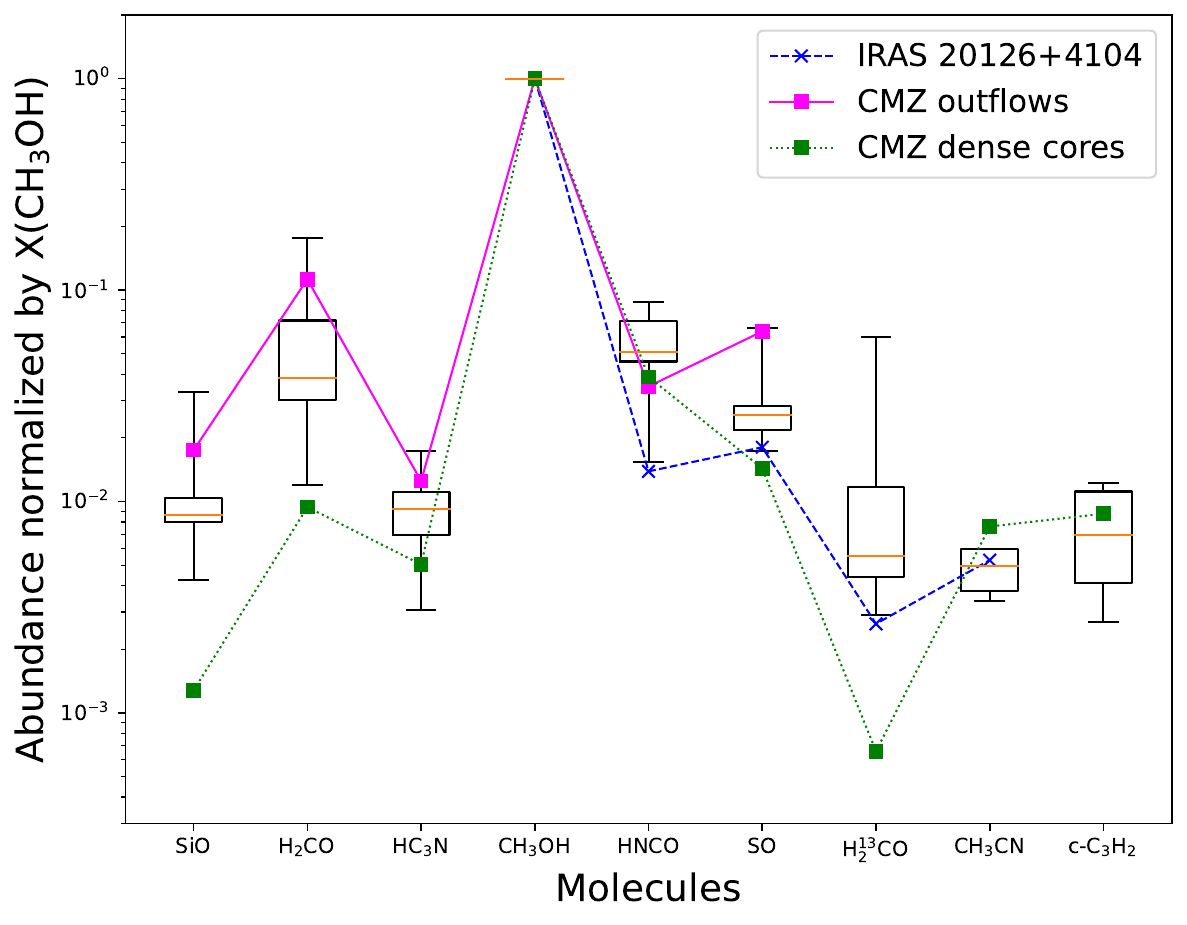}
\caption{Molecular abundances normalized with respected to the abundance of CH$_{3}$OH. The boxes denote the first to third quartiles while the caps mark the full range of abundances in our slim filaments. The median of abundances of each molecule is marked by a horizontal orange line. The abundances of the outflow from a high-mass protostar IRAS~20126+4104 \citep{2017MNRAS.467.2723P}, and of the dense cores \citepalias{lu2020} and outflows \citepalias{lu2021} in the CMZ are also plotted. The systematic uncertainties in the abundances are not plotted here.}
\label{fig:abund}
\end{figure}

We normalized the abundances of the nine molecules with respect to that of \meth{} and plot the relative abundances in \autoref{fig:abund}. To understand the relative abundances, we compared our results with outflow studies toward the same star-forming clouds in the CMZ \citepalias{lu2021}. Additionally, \mthc{} and \httco{} are rarely found in outflows. The outflow of a high-mass protostar IRAS~20126+4104 \citep{2017MNRAS.467.2723P} allows us to compare the column densities of these two molecules to that of \meth{}. The mean values of relative abundances in the outflows \citepalias{lu2021} and dense cores \citepalias{lu2020} in the CMZ, and IRAS 20126+4104 are also shown.

We performed a two-sample Kolmogorov-Smirnov test to compare the relative abundances in the slim filaments and outflows in the CMZ. In the test, a small \textit{p}-value (typically $<$ 0.05) indicates that the two distributions under consideration are significantly different. The \textit{p}-value is 0.68, indicating no significant evidence that they come from different distributions. We found no significant difference between the relative abundances in the filaments and the IRAS~20126+4104 outflow with a $p$-value of 0.14. However, the \textit{p}-value between the relative abundances in the filaments and dense cores in the CMZ is 0.02, indicating that they are statistically different from each other.

\section{Discussion}\label{sec:disc}

\subsection{A new class of filaments}\label{subsec:equi}

Many studies of filaments in nearby clouds are based on the assumption of hydrostatic equilibrium \citep[e.g.,][]{2011A&A...529L...6A,2016A&A...586A..27K,2018A&A...610A..77H}, which suggest a balance among pressures from turbulence, gravity, thermal motions, magnetic fields, and the external environment. In this study, we estimate these pressures and examine the radial equilibrium of the slim filaments.

The turbulent pressure can be estimated using $P_{\rm turb}$ = $\rho\sigma_{v}^{2}$. We adopt a mean $\sigma_{v}$ value of 4.4~\kms{} from Section \ref{subsec:linewidths}. The gas density, $\rho$ = $n m_{\rm H}$, depends on the density, $n$, which we estimate as follows. Given the slim filaments are detected with dust emission under a 5$\sigma$ level, the upper limit of the column density, $N_{\rm H_{2}}$, can be derived from the continuum emission using the following equation:
\begin{equation}\label{equa_col}
N_{\rm H_{2}} = \eta \frac{S_{\nu}}{B_{\nu}(T_{\rm dust})~\Omega~\kappa_{\nu}~\mu_{\rm H_{2}}~m_{p}}
\end{equation}
where $\eta$ = 100 is the assumed gas-to-dust mass ratio, $S_{\nu}$ is the continuum flux density, $B_{\nu}(T_{\rm dust})$ is the Plank function at dust temperature $T_{\rm dust}$ = 20~K following \citet{kauffmann2017a} and frequency $\nu$ = 226~GHz, $\Omega$ is the solid angle, $\kappa_{\nu}$ is the dust opacity, and $\mu_{\rm H_{2}}$ = 2.8 is the mean molecular weight of the interstellar medium \citep{2008A&A...487..993K}. We adopt $\kappa_{\nu}$ = 0.817 cm$^{2}$g$^{-1}$, assuming $\kappa_{\nu}$ = 10 $\times$ ($\nu$/1.2~THz)$^{\beta}$ cm$^{2}$g$^{-1}$ with $\beta$ = 1.5 \citep{1983QJRAS..24..267H}. Considering the dust emission at the 5$\sigma$ level of 0.2~\mjypbm{}, the upper limit for the column density is calculated to be 2.3 $\times$ 10$^{23}$~\sqc{}. Assuming that the thickness along the line of sight is the same as the slim filament width of 0.024~pc (see Section \ref{subsec:disc_width}), the estimated upper limit of the density $n$ is approximately 3.1 $\times$ 10$^{6}$ cm$^{-3}$. This yields a turbulent pressure $P_{\rm turb}$/$k_{\rm B}$ of approximately 1.7 $\times$ $10^{10}$~K\,\cc{}. 

The thermal pressure is calculated as $P_{\rm th}$/$k_{\rm B}$ = $\rho T$/$\mu_{p} m_{p}$, resulting in 9.2 $\times$ $10^{7}$~K\,\cc{}. The gravitational pressure, given by $P_{\rm gravity}$ $\propto$ $G\rho^{2}L^{2}$/$\pi$, is estimated to be $P_{\rm gravity}$/$k_{\rm B}$ of 1.3 $\times$ 10$^{8}$~K\,\cc{}. The magnetic pressure $P_{\rm mag}$/$k_{\rm B}$, which reaches the maximum $P_{\rm mag}$ = $B^{2}$/8$\pi$ when the magnetic field lines are parallel to the filament, is about 6.6 $\times$ $10^{7}$~K\,\cc{} for a magnetic strength of 0.48~mG from \citet{2024ApJ...962...39L}. Additionally, the external pressure, $P_{\rm ex}$/$k_{\rm B}$, is estimated using $\rho_{\rm ex} \sigma_{\rm ex}^{2}$ + $\rho_{\rm ex} T$/$\mu_{p} m_{p}$. We obtain median values from \cite{lu2019a}, with an external density of 4.4 $\times$ 10$^{5}$ cm$^{-3}$ and a external velocity dispersion of 1.43 \kms{}. Therefore, the external pressure is calculated to be approximately 2.6 $\times$ 10$^{8}$~K\,\cc{}.

Our results show that the turbulent pressure is two to three orders of magnitude higher than the other estimated pressures, suggesting that
the slim filaments are not in hydrostatic equilibrium. The high turbulent pressure could lead to the expansion and eventual dissipation of these filaments. These findings indicate that the slim filaments differ significantly from the dense gas filaments typically observed in nearby molecular clouds that are usually considered to be in hydrostatic equilibrium \citep[e.g.,][]{wang2014,lu2018}. 

With the filament width of 0.024~pc and the typical velocity dispersion of 4.5~\kms{}, the dissipation timescale of the filaments is estimated to be $\sim$5.2 $\times$ 10$^{3}$ years, which is comparable with the freeze-out timescale of SiO \citep[approximately 1 $\times$ 10$^{4}$ years,][]{1998ApJ...499..777B,1999A&A...343..585C}. Under the assumptions of LTE conditions, optically thin line emission, and an excitation temperature of 70~K, we estimate a SiO mass of 1.3~$M_{\odot}$ in the CMZ based on the SiO\,5--4 line emission within a velocity range of $-$50 to 120~\kms{} from APEX observations \citep{ginsburg2016}. This estimate covers an area of 325 $\times$ 56~pc$^{2}$ in the CMZ, with detailed calculations provided in \autoref{appd_sec:column}. This corresponds to a SiO depletion rate of $\sim$1.3 $\times$ 10$^{-4}$~$M_{\odot}$\,yr$^{-1}$.

For our detected slim filaments, the median SiO mass is 8.1 $\times$ 10$^{-5}$~$M_{\odot}$, meaning that each filament contributes an SiO replenishment rate of $\sim$1.6 $\times$ 10$^{-8}$~$M_{\odot}$\,yr$^{-1}$. To maintain a balance between the replenishment and depletion, about 8000 filaments would be required over the CMZ, which amounts to a slim filament surface density of $\sim$0.4~pc$^{-2}$. Toward the observed regions in \ctw{} and \cfi{} where the total area is $\sim$15~pc$^{-2}$, 10 slim filaments have been identified, leading to a surface density of $\sim$0.7~pc$^{-2}$.

If the slim filaments ubiquitously exist throughout the CMZ at a similar surface density, their dissipation would be sufficient to refuel the widespread SiO emission in the CMZ while this molecule is simultaneously freezing out onto dust grains, achieving a balance between the replenishment and depletion. Meanwhile, several COMs, such as \meth{}, \mthc{}, \fmh{}, and HC$_3$N in the slim filaments would be released into the interstellar medium (ISM), which might explain the widespread emission of COMs in the CMZ. A CMZ-wide census of slim filaments is necessary to confirm this possibility.

\subsection{Possible origin of the slim filaments}\label{subsubsec:disc_origin}

The slim filaments exhibit unique morphology, velocity structures, relative molecular abundances, and dynamic states, suggesting that they may have a different origin than the dense gas filaments found in nearby clouds. We find several clues suggesting that the slim filaments are related to shock activities:
\begin{itemize}
\item The rotational transitions of SiO in the ISM are usually suggested to trace shocks because Si atoms can be released from dust grains through sputtering or vaporization caused by shock activities \citep{schilke1997}. The shock activities consist of high-velocity shocks from outflows and low-velocity shocks with unclear origins \citep[e.g.,][]{2014A&A...570A...1D,2016A&A...595A.122L,2024ApJ...960...48T}.

\item Class~\textsc{i} \meth{} masers are believed to be collisionally pumped \citep{2014MNRAS.439.2584V}, and are found in all the four regions with slim filaments (marked by orange crosses in \autoref{fig:maser}). While widespread \meth{} masers are observed in the CMZ and are thought to result from photodesorption driven by cosmic rays \citep{yusefzadeh2013}, the masers in these regions are likely generated by shock activities, as their velocities are consistent with the $V_{\rm lsr}$ of the SiO emissions observed in the filaments, hinting a shocked environment.

\item Statistically, the relative abundances of the molecules in the slim filaments are similar to those in other shocked regions such as protostellar outflows, while they are different from those in dense cores, suggesting their closer relation to shocks instead of protostellar heating.
\end{itemize}

Under shock conditions, the non-detection of thermal dust emission at the 5$\sigma$ level can be explained if: (i) the shocks have destroyed most, if not all, of the dust grains, releasing Si into the gas phase; or (ii) the initial dust emission is diffuse, and a shock wave can efficiently sputter Si atoms and form SiO. In the latter case, the dust emission remains undetectable by ALMA due to missing flux, a phenomenon similar to shocks in outflows interacting with surrounding gas.

In \autoref{fig:zoomin}, we notice that the regions containing slim filaments are located at the edges of \ctw{} and \cfi{}, where they are exposed to and can more easily interact with the external environment. 
Additionally, \autoref{fig:maser} presents the magnetic orientations with a resolution of 19.6$^{\prime\prime}$ obtained from \cite{2024ApJ...969..150P}. We find that magnetic fields are nearly perpendicular to the main filament skeletons. 
Studies modeling the influence of a shock wave on a molecular cloud suggest that shock compression can lead to the formation of filaments oriented perpendicular to the magnetic field \citep[e.g.,][]{2018PASJ...70S..53I,2021ApJ...916...83A}.
Given the likely presence of shocks, we speculate that these filaments result from the interaction of shock waves with magnetized molecular clouds. 
Future multi-wavelength investigations that better constrain excitation conditions and chemistry will help confirm the nature of these filaments.

\section{Conclusions}\label{sec:conclusions}
We have identified slim filaments in four regions toward the CMZ through ALMA 1.3~mm molecular lines. These filaments are distinctive for their narrow ($\lesssim$0.03~pc) morphology, with strong SiO 5--4 line emissions as well as non-detections of thermal dust emission at the 5$\sigma$ level. They show consistent velocity structures which are different from outflows. The analysis of various pressures in slim filaments suggests that turbulent pressure dominates, leading to hydrostatic inequilibrium and therefore potential expansion and dissipation. The relative molecular abundances are statistically similar with those in protostellar outflows and the detection of collisionally pumped Class \textsc{i} \meth{} masers hints an association with shocks.

We speculate that these slim filaments represent a distinct class from the dense gas filaments typically observed in nearby molecular clouds, and they may result from interactions between shocks and molecular clouds. 
Their eventual dissipation within $\sim$10$^4$ years may enrich SiO and several COMs (e.g., \meth{}, \mthc{}, \fmh{}, HC$_3$N) in the ISM, thus leading to the observed widespread emission of SiO and COMs in the CMZ.

\begin{acknowledgements}
We thank Di Li, Dalei Li, and Natalie Butterfield for helpful discussions.
This work is supported by the National Key R\&D Program of China (No.\ 2022YFA1603100) and the Strategic Priority Research Program of the Chinese Academy of Sciences (CAS) Grant No.\ XDB0800300. 
K.Y. acknowledges support from the Shanghai Post-doctoral Excellence Program (No. 2024379).
X.L.\ acknowledges support from the National Natural Science Foundation of China (NSFC) through grant Nos.\ 12273090 and 12322305, the Natural Science Foundation of Shanghai (No.\ 23ZR1482100), and the CAS ``Light of West China'' Program No.\ xbzg-zdsys-202212.
Y.Z. acknowledges the support from the Yangyang Development Fund.
H.B.L.\ is supported by the National Science and Technology Council (NSTC) of Taiwan (Grant Nos.\ 111-2112-M-110-022-MY3, 113-2112-M-110-022-MY3).
Q.Z.\ acknowledges the support of National Science Foundation under award No.\ 2206512.
E.A.C.\ Mills gratefully acknowledges funding from the National Science Foundation under Award Nos.\ 1813765, 2115428, 2206509, and CAREER 2339670.
JMDK gratefully acknowledges funding from the European Research Council (ERC) under the European Union's Horizon 2020 research and innovation programme via the ERC Starting Grant MUSTANG (grant agreement number 714907). 
COOL Research DAO is a Decentralized Autonomous Organization supporting research in astrophysics aimed at uncovering our cosmic origins.
S.F. acknowledges support by the National Science Foundation of China (12373023, 12133008). AG acknowledges support from the NSF under grants AAG 2008101, 2206511, and CAREER 2142300.
H.-H.Q.\ is partially supported by the Youth Innovation Promotion Association CAS and the NSFC (grant No.\ 11903038) and the Strategic Priority Research Program of the Chinese Academy of Sciences (grant No.\ XDB1070301 and XDB1070302).
This paper makes use of the following ALMA data: ADS/JAO.ALMA\#2016.1.00243.S, ADS/JAO.ALMA\#2016.1.00875.S\@. ALMA is a partnership of ESO (representing its member states), NSF (USA) and NINS (Japan), together with NRC (Canada), MOST and ASIAA (Taiwan), and KASI (Republic of Korea), in cooperation with the Republic of Chile. The Joint ALMA Observatory is operated by ESO, AUI/NRAO and NAOJ\@. Data analysis was in part carried out on the open-use data analysis computer system at the Astronomy Data Center (ADC) of NAOJ\@. This research has made use of NASA’s Astrophysics Data System. This research has made use of CASA \citep{2022PASP..134k4501C}, APLpy \citep{aplpy2012}, Astropy \citep{astropy2013}, and pvextractor(\url{http://pvextractor.readthedocs.io}).

\end{acknowledgements}

\bibliographystyle{aa}
\bibliography{my_2024_cmzfilaments}

\begin{appendix}

\onecolumn

\section{Integrated maps of different molecular lines}\label{app_sec:fila}

\begin{figure*}
\centering
\includegraphics[width=0.9\textwidth]{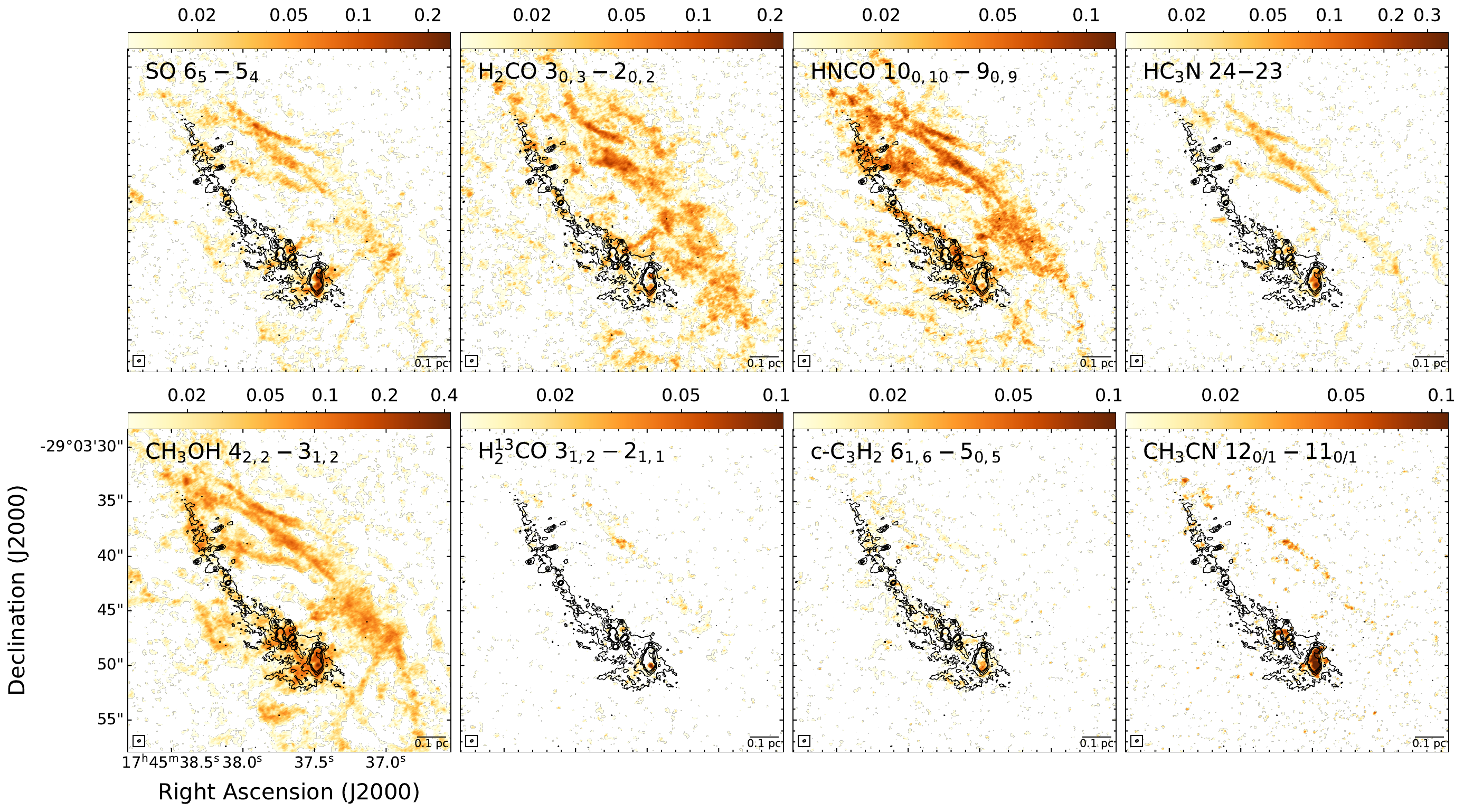}
\caption{Integrated maps of molecules in \autoref{tab_lines} for the slim filaments in the \ctw-A region.}
\label{fig:m0_a_app}
\end{figure*}

\begin{figure*}
\centering
\includegraphics[width=0.9\textwidth]{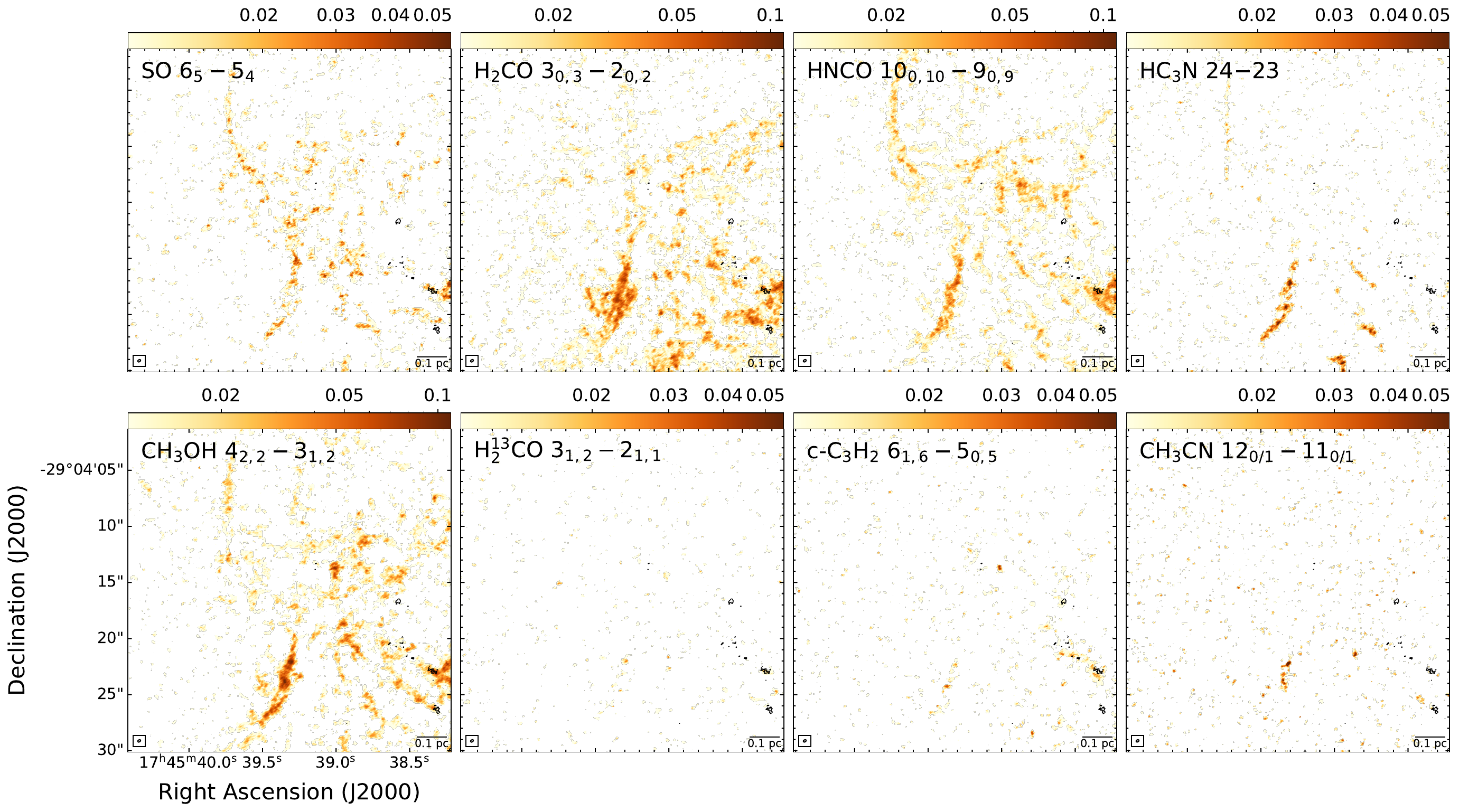}
\caption{Integrated maps of molecules in \autoref{tab_lines} for the slim filaments in the \ctw-B region..}
\label{fig:m0_b_app}
\end{figure*}

\begin{figure*}
\centering
\includegraphics[width=0.9\textwidth]{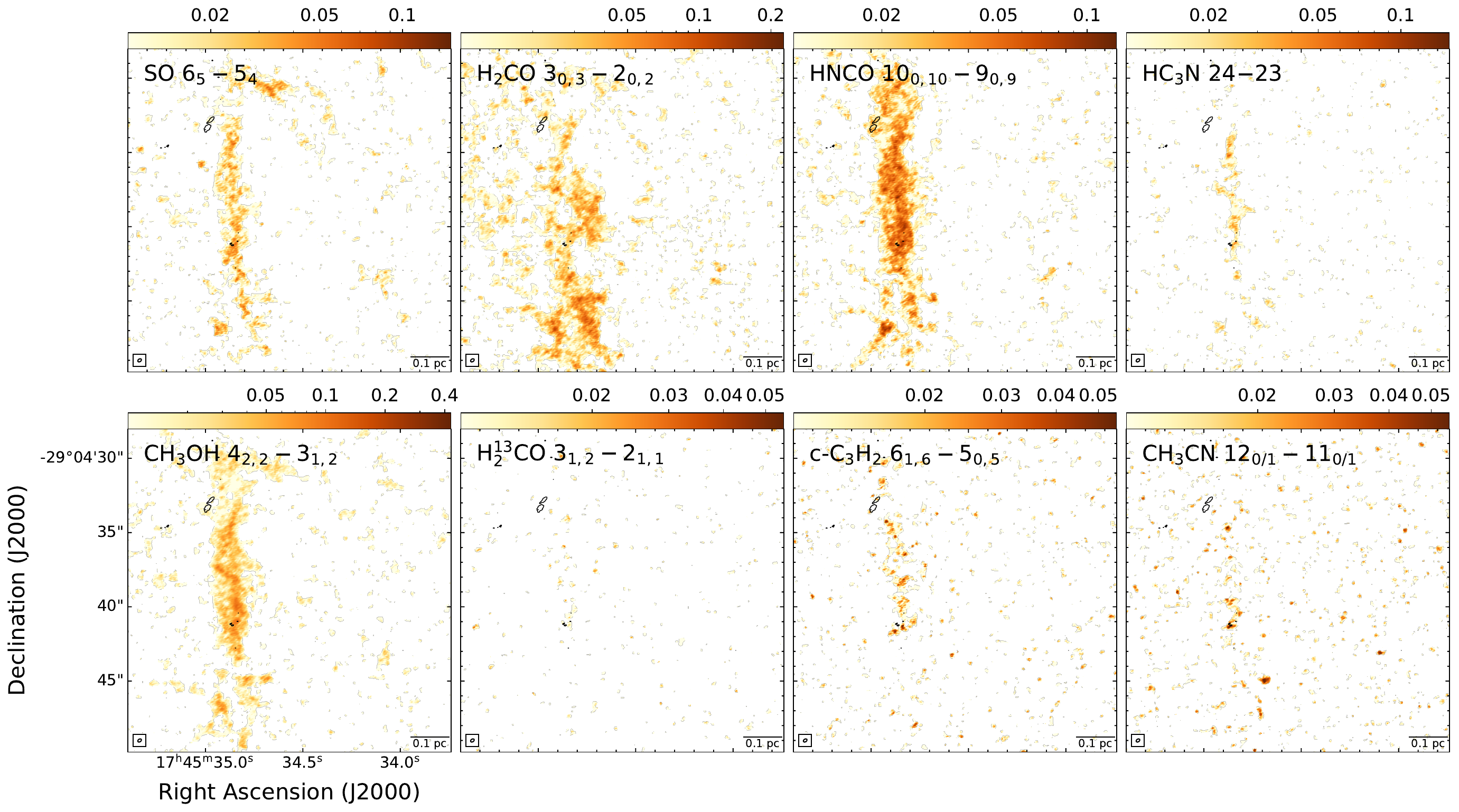}
\caption{Integrated maps of molecules in \autoref{tab_lines} for the slim filaments in the \ctw-C region..}
\label{fig:m0_c_app}
\end{figure*}

The integrated maps of SO 6$_{5}$--5$_{4}$, \fmh{} 3$_{0,3}$--2$_{0,2}$, HNCO 10$_{0,10}$--9$_{0,9}$, \cyacet{} 24--23, \meth{} 4$_{2,2}$--3$_{1,2}$, \httco{} 3$_{1,2}$--2$_{1,1}$, \cctht{} 6$_{1,6}$--5$_{0,5}$, and \mthc{} 12$_{0/1}$--11$_{0/1}$ for the slim filaments in \ctw{} are presented in Figures~\ref{fig:m0_a_app}--\ref{fig:m0_c_app}.

\section{Skeletons identified by FilFinder}\label{app_sec:filfinder}

\begin{figure*}
\centering
\includegraphics[width=0.46\textwidth]{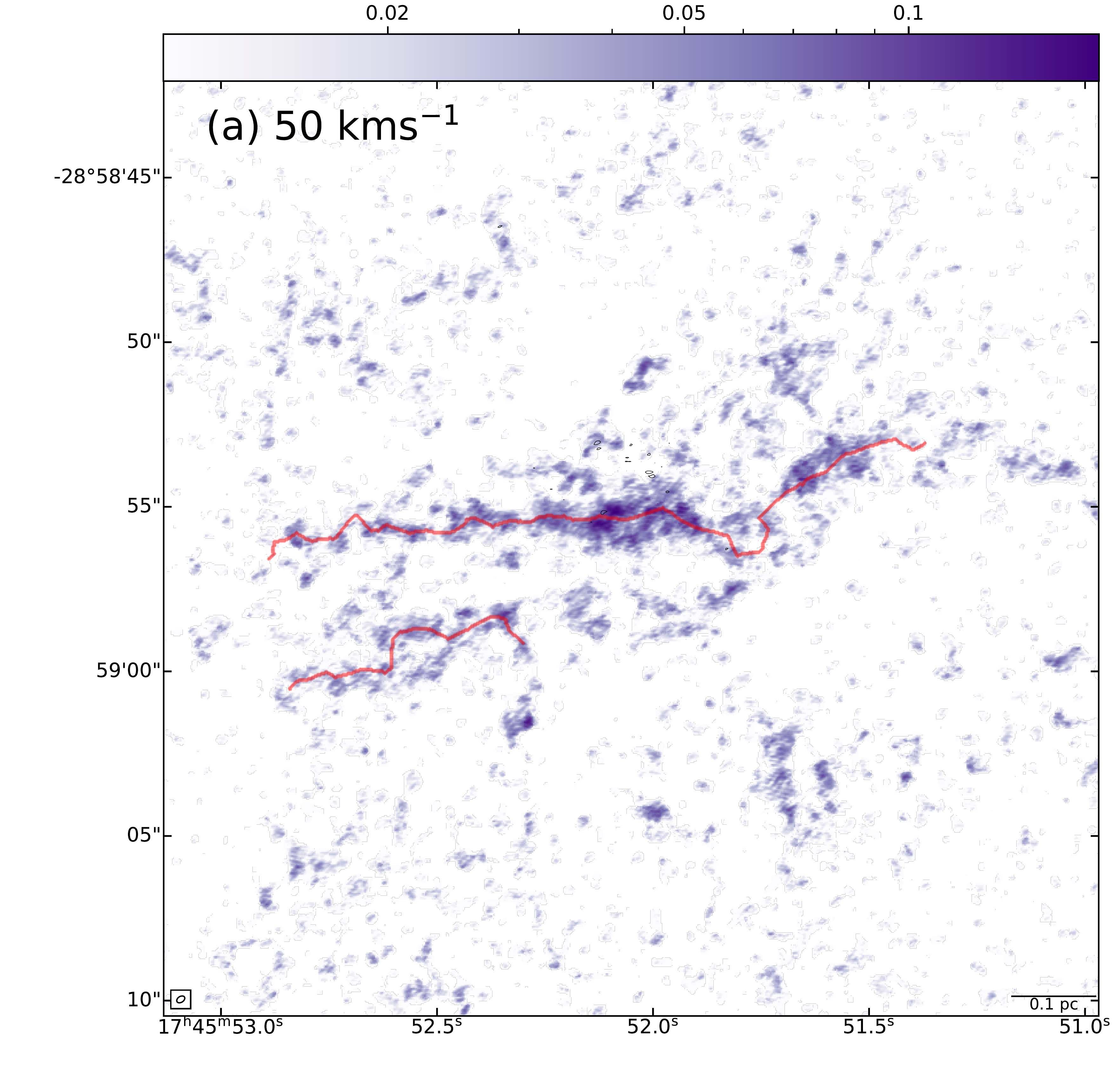}
\includegraphics[width=0.45\textwidth]{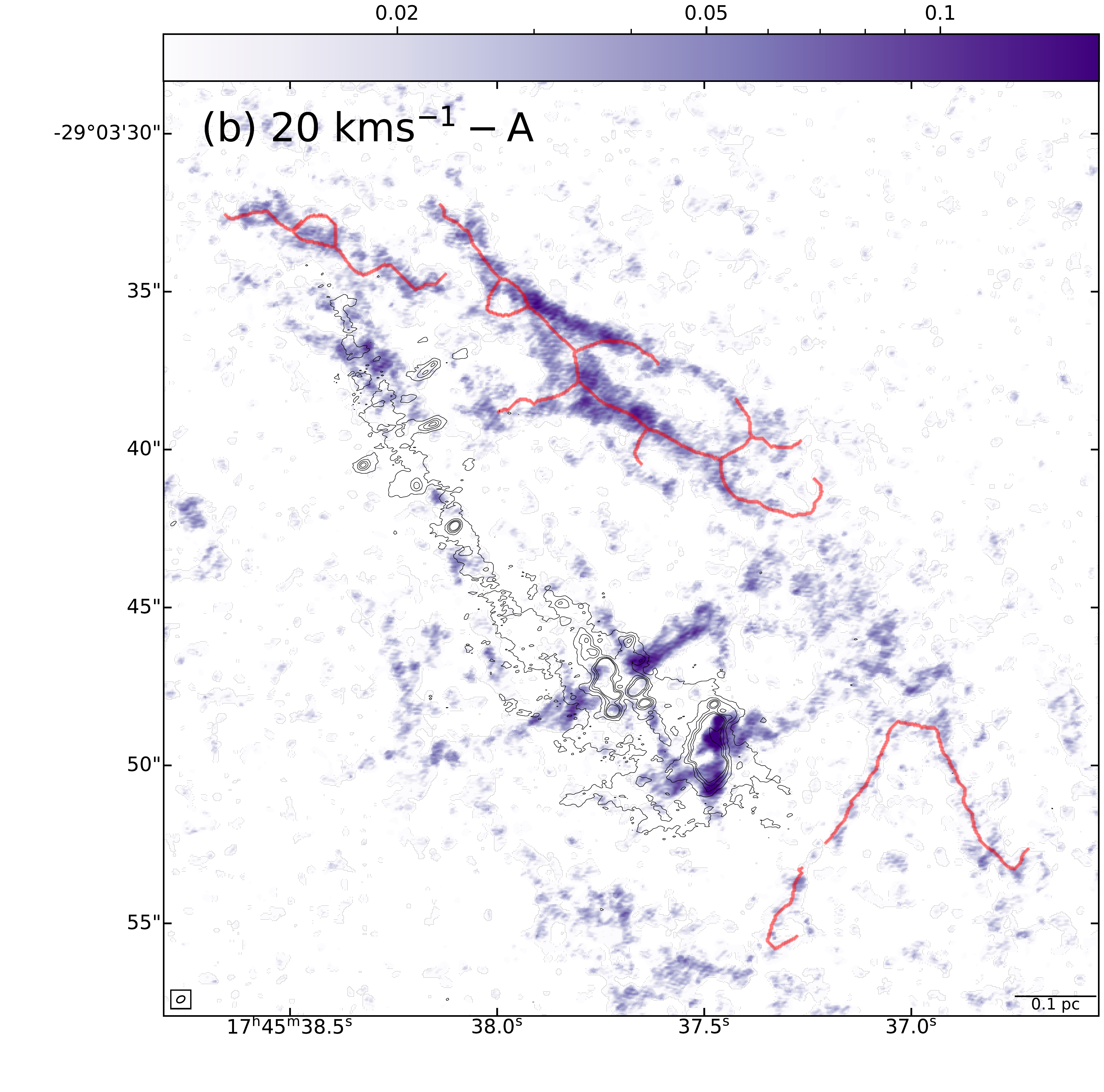}\\
\includegraphics[width=0.45\textwidth]{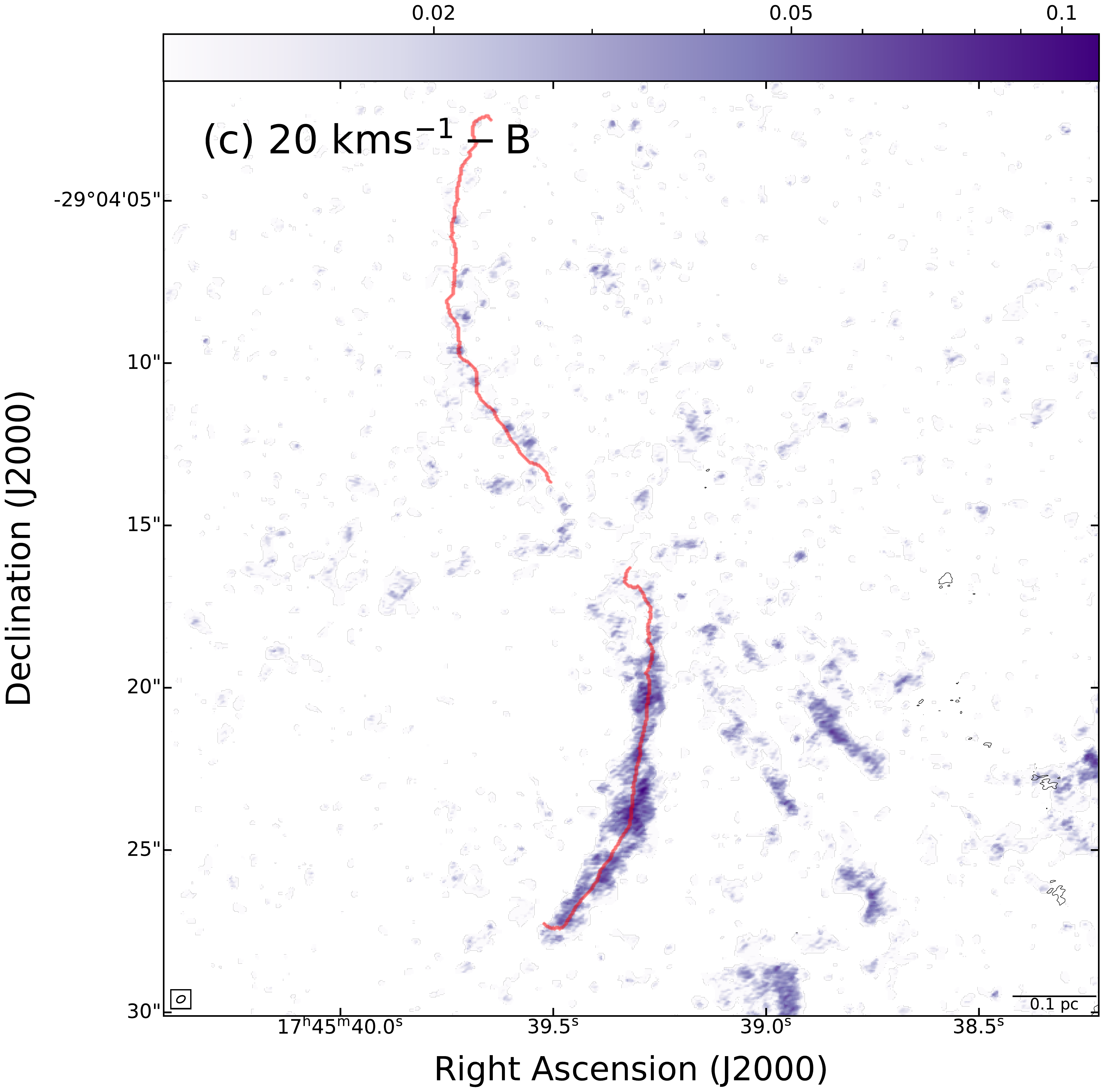}
\includegraphics[width=0.45\textwidth]{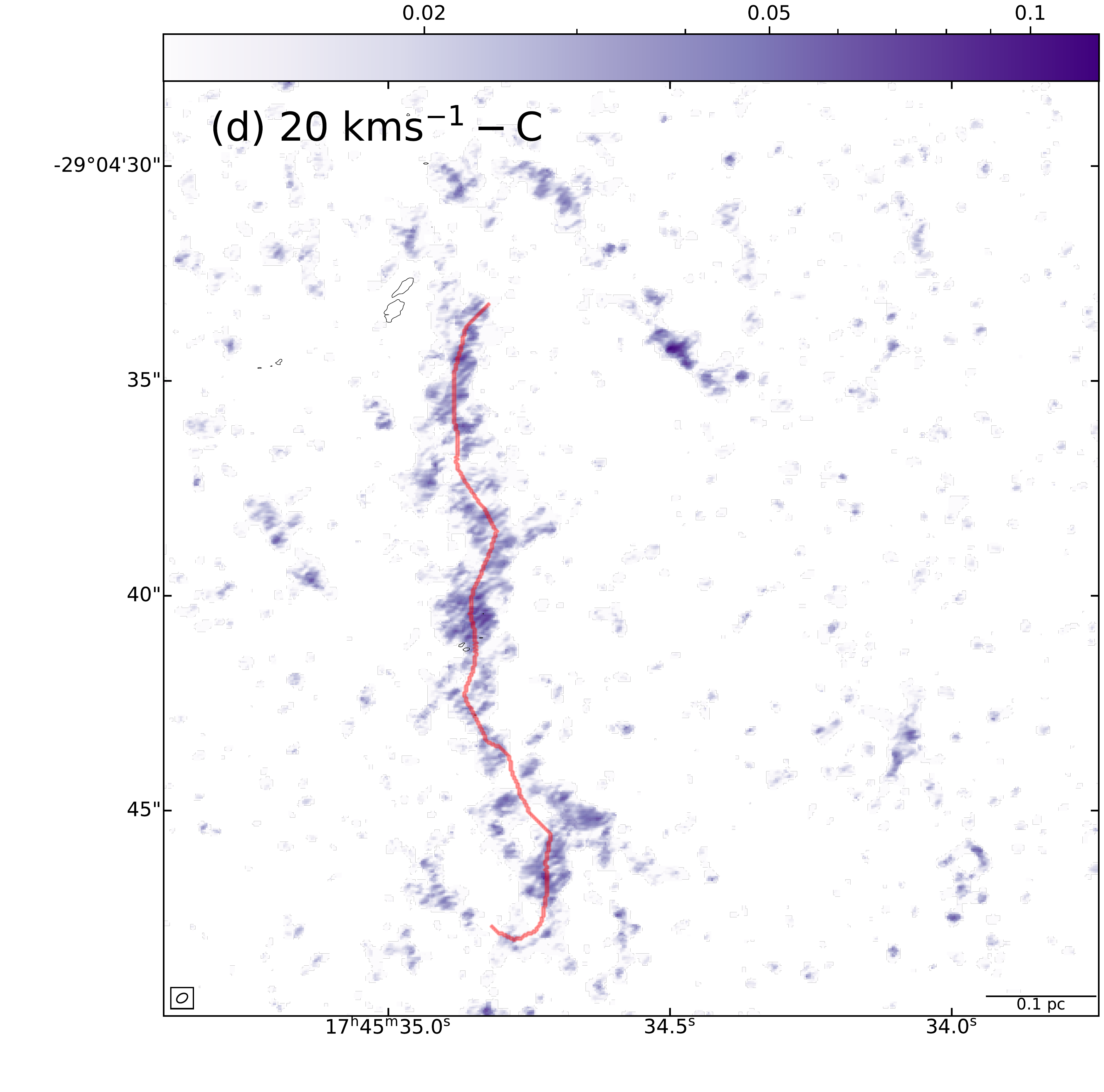}
\caption{SiO\,5--4 emission of filaments in four regions. The red lines indicate the skeletons of the slim filaments.}
\label{fig:skele}
\end{figure*}

The \texttt{FilFinder} package reduces the selecting area to identify skeletons that represent the topology of the areas, by using a Medial Axis Transform method. To achieve optimal filament detection, we set the following parameters: (i) global threshold - integrated intensities below 3$\sigma_{\rm area}$ are excluded from the mask; (ii) size threshold - filaments must have a minimum area of 125 pixel$^{2}$ ($\sim$ 5 times the beam size). The resulting skeletons are depicted as red lines in \autoref{fig:skele}.

\section{Gaussian fits to the SiO 5--4 spectra in four regions}\label{appd_sec:widths}

\begin{figure*}
\centering
\includegraphics[width=0.46\textwidth]{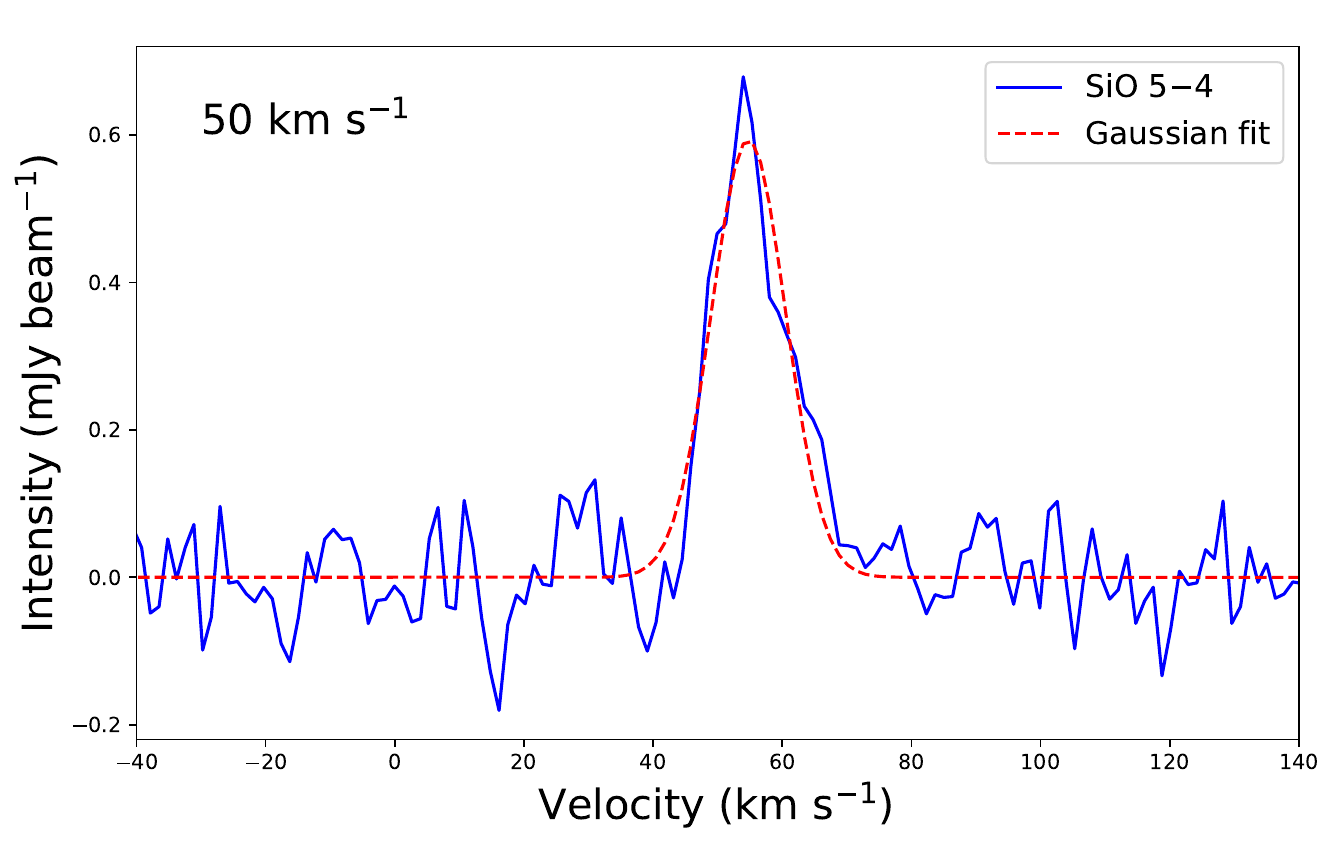}
\includegraphics[width=0.45\textwidth]{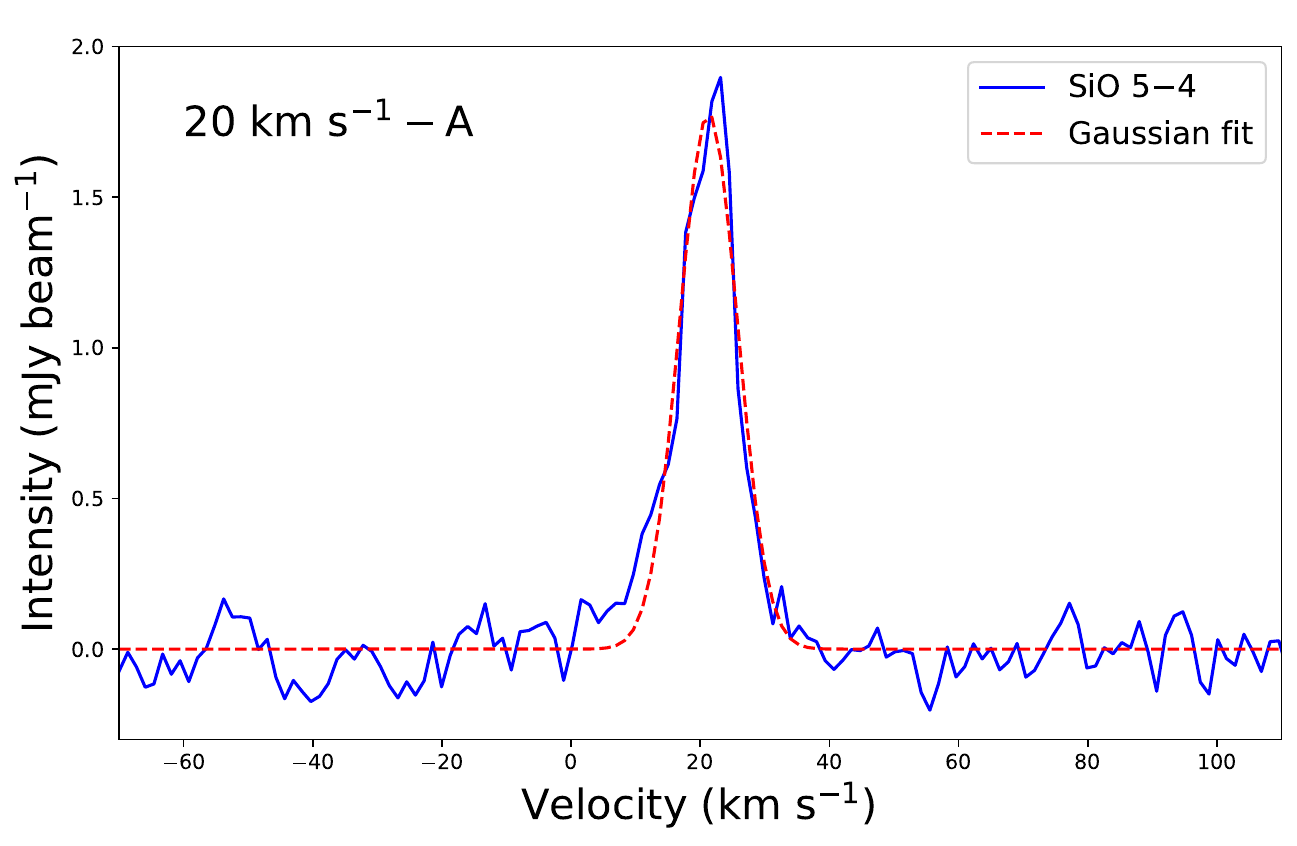}\\
\includegraphics[width=0.45\textwidth]{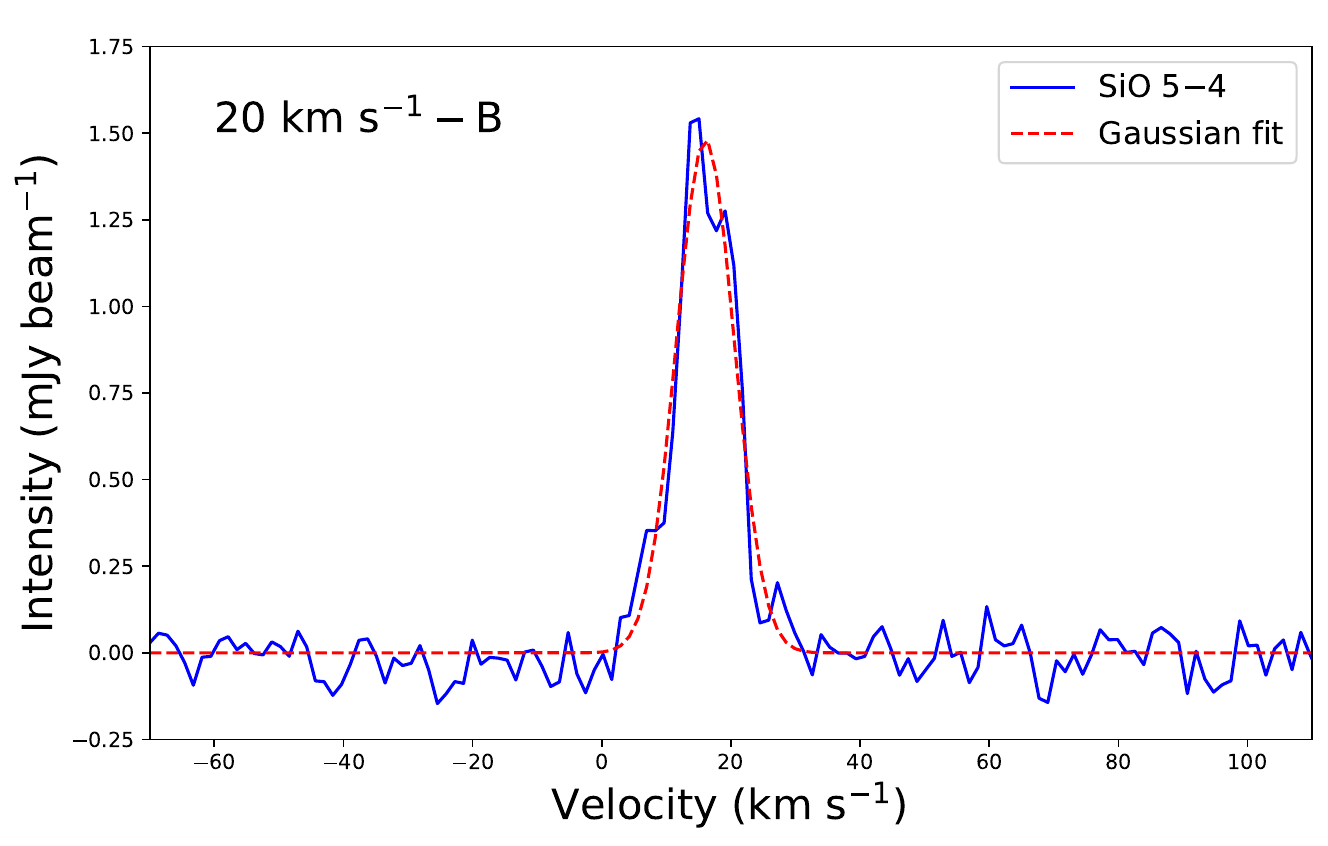}
\includegraphics[width=0.45\textwidth]{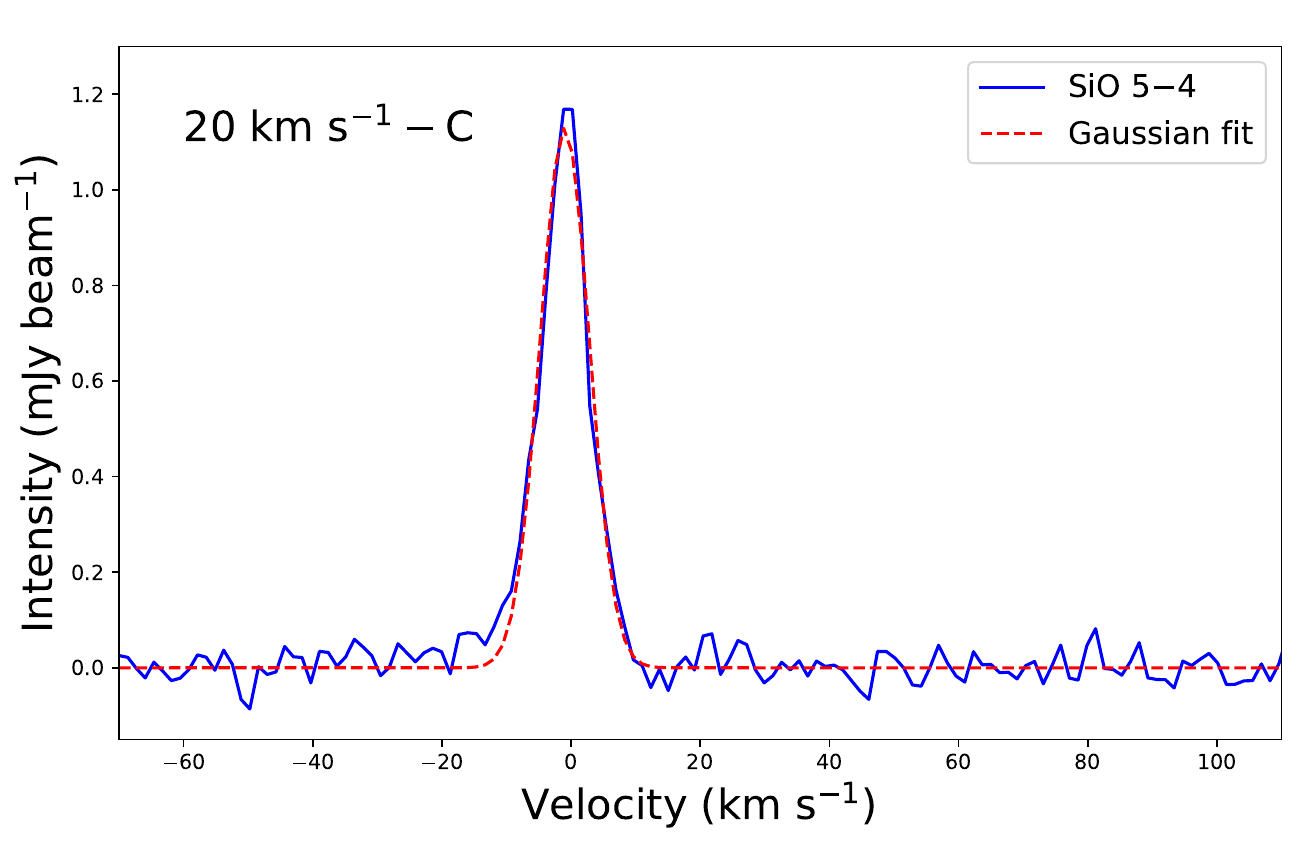}
\caption{Spectra observed in the SiO\,5--4 transition averaged over the filaments areas in four regions. The best Gaussian fits to the spectra are overlaid in pink.}
\label{fig:gauss}
\end{figure*}

We perform Gaussian fits to the averaged SiO 5--4 spectra toward the four regions with results shown in \autoref{fig:gauss}.

\section{Calculation of Molecular Column Densities}\label{appd_sec:column}

Assuming local thermodynamic equilibrium (LTE) conditions, optically thin line emission, Rayleigh-Jeans approximation, and negligible background, the column densities of a molecule can be derived following \citep{mangum2015}:
\begin{equation}\label{Eq_N}
N_{\rm tot} = \frac{8 \pi k_{\rm B} \nu^{2}}{h c^{3} A_{ul}} \frac{Q(T_{\rm ex})}{g_{u}} {\rm exp} \left(\frac{E_{u}}{k_{\rm B} T_{\rm ex}}\right) \int T_{B}dv, \\
\end{equation}
where $k_{\rm B}$ is the Boltzmann constant, $\nu$ is the rest frequency of the transition, $h$ is the Planck constant, $c$ is the light speed, $A_{ul}$ is the spontaneous emission coefficient from the upper state $u$ to the lower state $l$, $Q(T_{\rm ex})$ is the partition function, $T_{\rm ex}$ is the excitation temperature, $g_{u}$ is the degeneracy of the upper state, $E_{u}$ is the energy of the upper level energy, and $\int T_{B}dv$ is the integrated intensity. The spectroscopic parameters are obtained from the CDMS database \citep{2001A&A...370L..49M,2005JMoSt.742..215M,2016JMoSp.327...95E}, JPL catalogues \citep{1998JQSRT..60..883P}, and the LAMDA database \citep{lamda2005}. The excitation temperature is adopted to be 70~K following \citetalias{lu2021}. The derived column densities are presented in \autoref{tab_column}.

The SiO mass, $M_{\rm SiO}$, can estimated based on the SiO column density \citep{goldsmith1999}:
\begin{equation}\label{Eq_mass}
M_{\rm SiO} = N_{\rm tot} \mu_{\rm SiO}m_{p}\Omega d^{2}, \\
\end{equation}
where $\mu_{\rm SiO}$ is the mean atomic weight of 44 for SiO, $\Omega$ is the solid angle, and $d$ is the distance of 8.1 kpc.

\clearpage

\begin{sidewaystable*}
\caption{Molecular column densities toward SiO peak positions.} \label{tab_column}
\begin{center}
\begin{tabular}{c c c cc c ccccccccc}
\hline\hline
 Cloud & Region & & \multicolumn{2}{c}{Coordinates} & & \multicolumn{9}{c}{Column densities} \\
 \cline{4-5} \cline{7-15}
 & & & R.A. & Decl. & & SiO & SO & \meth{} & \fmh{} & \cyacet{} & HNCO & \httco{} & \cctht{} & \mthc{} \\
 & & & (hh:mm:ss) & (dd:mm:ss) & & (cm$^{-2}$) & (cm$^{-2}$) & (cm$^{-2}$) & (cm$^{-2}$) & (cm$^{-2}$) & (cm$^{-2}$) & (cm$^{-2}$) & (cm$^{-2}$) & (cm$^{-2}$) \\
\hline
50 km s$^{-1}$ & & & 17:45:52.087 & $-$28:58:55.150 & & 1.54$\times$10$^{14}$ & 4.57$\times$10$^{14}$ & 1.83$\times$10$^{16}$ & 2.86$\times$10$^{14}$ & 1.86$\times$10$^{14}$ & 2.82$\times$10$^{14}$ & --- & 4.91$\times$10$^{13}$ & 1.09$\times$10$^{14}$ \\
 & & & 17:45:52.340 & $-$28:58:58.224 & & 9.50$\times$10$^{13}$ & 1.70$\times$10$^{14}$ & 9.54$\times$10$^{15}$ & 3.60$\times$10$^{14}$ & 6.25$\times$10$^{13}$ & 2.05$\times$10$^{14}$ & --- & --- & --- \\
\hline
20 km s$^{-1}$ & A & & 17:45:37.906 & $-$29:03:35.371 & & 1.37$\times$10$^{14}$ & 6.28$\times$10$^{14}$ & 9.49$\times$10$^{15}$ & 1.27$\times$10$^{15}$ & 1.65$\times$10$^{14}$ & 4.63$\times$10$^{14}$ & 1.27$\times$10$^{14}$ & 9.02$\times$10$^{13}$ & 3.72$\times$10$^{13}$ \\
 & & & 17:45:38.540 & $-$29:03:32.649 & & 6.47$\times$10$^{13}$ & 1.51$\times$10$^{14}$ & 8.67$\times$10$^{15}$ & 1.04$\times$10$^{14}$ & 8.03$\times$10$^{13}$ & 4.05$\times$10$^{14}$ & 4.77$\times$10$^{13}$ & --- & --- \\
 & & & 17:45:37.664 & $-$29:03:38.772 & & 1.09$\times$10$^{14}$ & 3.17$\times$10$^{14}$ & 1.26$\times$10$^{16}$ & 9.20$\times$10$^{14}$ & 1.31$\times$10$^{14}$ & 8.91$\times$10$^{14}$ & 1.27$\times$10$^{14}$ & 5.49$\times$10$^{13}$ & 7.49$\times$10$^{13}$ \\
 & & & 17:45:38.030 & $-$29:03:38.573 & & 5.48$\times$10$^{13}$ & 1.00$\times$10$^{14}$ & 5.36$\times$10$^{15}$ & 3.78$\times$10$^{14}$ & 6.29$\times$10$^{13}$ & 4.70$\times$10$^{14}$ & --- & 2.17$\times$10$^{13}$ & --- \\
 & & & 17:45:36.996 & $-$29:03:47.614 & & 7.16$\times$10$^{13}$ & 2.55$\times$10$^{14}$ & 9.41$\times$10$^{15}$ & 3.62$\times$10$^{14}$ & 2.89$\times$10$^{13}$ & 4.25$\times$10$^{14}$ & 2.56$\times$10$^{13}$ & --- & --- \\
 & & & 17:45:36.831 & $-$29:03:53.011 & & 5.69$\times$10$^{13}$ & 7.79$\times$10$^{13}$ & 1.73$\times$10$^{15}$ & 3.05$\times$10$^{14}$ & 1.26$\times$10$^{13}$ & 8.84$\times$10$^{13}$ & 2.74$\times$10$^{13}$ & --- & --- \\
\cline{2-15}
 & B & & 17:45:39.293 & $-$29:04:22.061 & & 6.17$\times$10$^{13}$ & 2.09$\times$10$^{14}$ & 7.15$\times$10$^{15}$ & 4.53$\times$10$^{14}$ & 9.70$\times$10$^{13}$ & 5.16$\times$10$^{14}$ & 1.04$\times$10$^{14}$ & 8.37$\times$10$^{13}$ & --- \\ 
 & & & 17:45:39.576 & $-$29:04:12.630 & & 3.02$\times$10$^{13}$ & 7.79$\times$10$^{13}$ & 2.87$\times$10$^{15}$ & 8.74$\times$10$^{13}$ & 1.22$\times$10$^{13}$ & 1.73$\times$10$^{14}$ & --- & --- & --- \\
\cline{2-15}
 & C & & 17:45:34.835 & $-$29:04:40.047 & & 4.27$\times$10$^{13}$ & 2.56$\times$10$^{14}$ & 1.00$\times$10$^{16}$ & 2.98$\times$10$^{14}$ & 9.25$\times$10$^{13}$ & 7.17$\times$10$^{14}$ & 2.87$\times$10$^{13}$ & 1.22$\times$10$^{14}$ & 3.38$\times$10$^{13}$ \\
\hline\hline
\end{tabular}
\end{center}
\end{sidewaystable*}

\end{appendix}

\end{document}